\pdfoutput=1

\documentclass[11pt,twoside,a4paper,cmspaper,final,collab]{cms-tdr}
\def\svnVersion{485433P}\def\cmsCernNoTag{CERN-EP-2018-125}\def\cmsCernDate{\today}\def\cmsMessage{Published in Physical Review D as \href{http://dx.doi.org/10.1103/PhysRevD.98.112011}{\doi{10.1103/PhysRevD.98.112011.}}}

\begin{document}\cmsNoteHeader{BPH-15-001}

\hyphenation{had-ron-i-za-tion}
\hyphenation{cal-or-i-me-ter}
\hyphenation{de-vices}
\RCS$HeadURL: svn+ssh://svn.cern.ch/reps/tdr2/papers/BPH-15-001/trunk/BPH-15-001.tex $
\RCS$Id: BPH-15-001.tex 485378 2018-12-21 02:13:02Z wangdy $

\newlength\cmsFigWidth
\newlength\cmsTabSkip\setlength{\cmsTabSkip}{1ex}
\providecommand{\cmsTable}[1]{\resizebox{\textwidth}{!}{#1}}
\ifthenelse{\boolean{cms@external}}{\setlength\cmsFigWidth{0.85\columnwidth}}{\setlength\cmsFigWidth{0.4\textwidth}}
\ifthenelse{\boolean{cms@external}}{\providecommand{\cmsLeft}{top\xspace}}{\providecommand{\cmsLeft}{left\xspace}}
\ifthenelse{\boolean{cms@external}}{\providecommand{\cmsRight}{bottom\xspace}}{\providecommand{\cmsRight}{right\xspace}}
\ifthenelse{\boolean{cms@external}}{\providecommand{\NA}{\ensuremath{\cdots}\xspace}}{\providecommand{\NA}{\ensuremath{\text{---}}\xspace}}
\providecommand{\cmsPX}{\ensuremath{\cmsSymbolFace{X}}\xspace}
\newcommand{\Sig}{\ensuremath{\PBp \to \PKp \Pgmp \Pgmm}\xspace}
\newcommand{\bts}{\ensuremath{\PQb \to \PQs \ell^+ \ell^-}\xspace}
\newcommand{\JPS}{\ensuremath{\PBp \to \PKp \JPsi (\Pgmp\Pgmm)}\xspace}
\newcommand{\PsP}{\ensuremath{\PBp \to \PKp \Pgy (\Pgmp\Pgmm)}\xspace}
\newcommand{\PSX}{\ensuremath{\PBp \to \Pgmp \Pgmm \cmsPX}\xspace}

\cmsNoteHeader{BPH-15-001}
\title{Angular analysis of the decay \texorpdfstring{\Sig}{B+ to K+ mu+ mu-} in proton-proton collisions at \texorpdfstring{$\sqrt{s} = 8\TeV$}{sqrt(s) = 8 TeV}}

\date{\today}

\abstract{The angular distribution of the flavor-changing neutral
  current decay \Sig is studied in proton-proton collisions at a
  center-of-mass energy of 8\TeV. The analysis is based on data
  collected with the CMS detector at the LHC, corresponding to an
  integrated luminosity of 20.5\fbinv. The forward-backward asymmetry
  $A_{\mathrm{FB}}$ of the dimuon system and the contribution
  $F_{\PH}$ from the pseudoscalar, scalar, and tensor
  amplitudes to the decay width are measured as a function of the
  dimuon mass squared.  The measurements are consistent with the
  standard model expectations.}

\hypersetup{
pdfauthor={CMS Collaboration},
pdftitle={Angular analysis of the decay B+ to K+ mu+ mu- in proton-proton collisions at sqrt[s] = 8 TeV},
pdfsubject={CMS},
pdfkeywords={CMS, physics, B+ decays, angular analysis}
}

\maketitle

\section{Introduction}
\label{sec:intro}

The decay \Sig is a manifestation of a flavor-changing neutral
current process of the type \bts, with $\ell$ denoting a charged
lepton. In the standard model (SM), this decay is forbidden at tree
level and occurs through higher-order processes. This makes the
measurement of this process more
sensitive to possible physics phenomena beyond the SM (BSM).

\begin{figure}[htbp]
    \centering
	\includegraphics[width=0.3\textwidth]{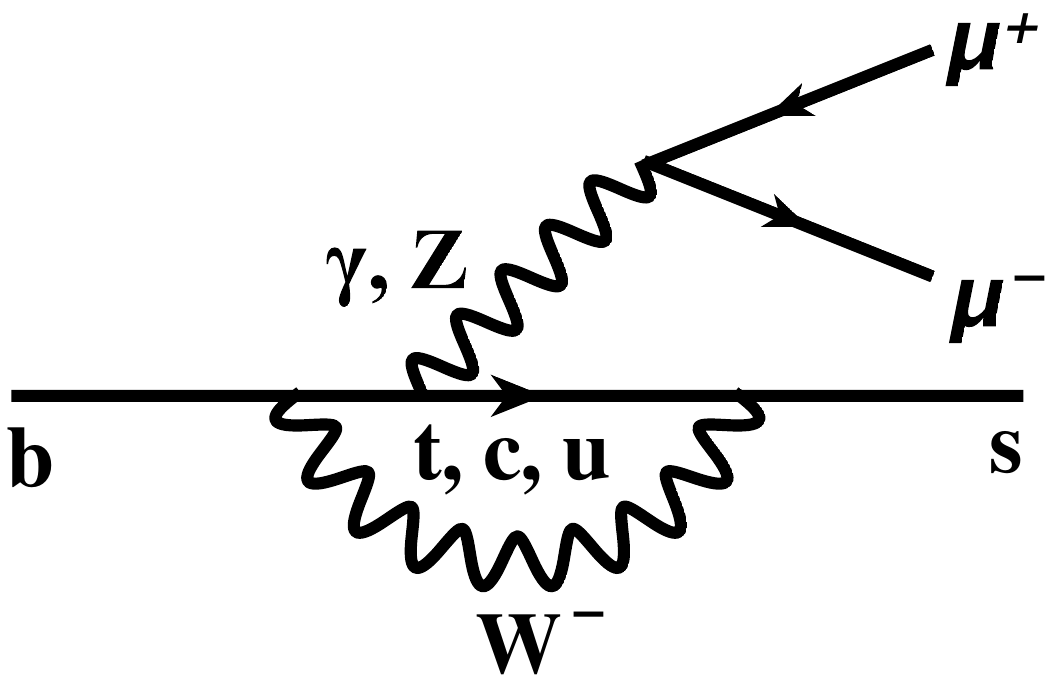}
	\includegraphics[width=0.3\textwidth]{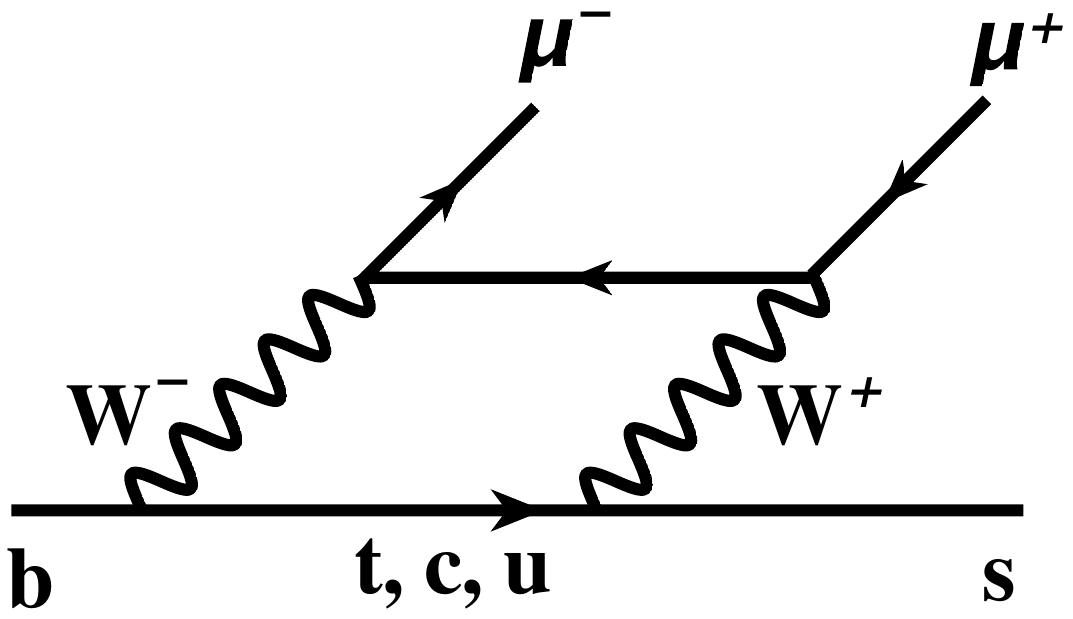}
        \caption{\label{fig-diagrams} The SM electroweak
          $\PZ/\Pgg$ penguin (\cmsLeft) and $\PWp\PWm$
          box (\cmsRight) diagrams for the decay process \Sig. }
\end{figure}

In the SM, three amplitudes contribute to \Sig via either electroweak
$\PZ/\Pgg$ penguin diagrams or a $\PWp\PWm$ box
diagram, as shown in Fig.~\ref{fig-diagrams}.
Two independent parameters describe the decay rate for the \Sig
process: the forward-backward asymmetry $A_{\mathrm{FB}}$ of the
dimuon system and the contribution $F_{\PH}$ from the
pseudoscalar, scalar, and tensor amplitudes to the decay
width. Theoretical predictions are available for both
parameters~\cite{Theory:2007,Khodjamirian2010,Bobeth:2011nj}. In the
SM, $A_{\mathrm{FB}}$ is zero up to small corrections, and
$F_{\PH}$ is also small. Because SM amplitudes may interfere
with the contributions from BSM particles in loop diagrams, the decay
can probe the presence of yet-unobserved particles and
processes~\cite{Theory:1991,Theory:2015,Altmannshofer2017,Altmannshofer2012,ref2000,ref2001}.
For example, a nonzero $A_{\mathrm{FB}}$ or large $F_{\PH}$
would point to a BSM contribution~\cite{Theory:2007, Alok:2008wp},
which can be probed~\cite{SALA2017205,Altmannshofer2015} by comparing
the experimental measurements with the theoretical
predictions~\cite{Altmannshofer2017,Alok:2008wp,Ghosh2014}.

In this paper, we report the measurement of $A_{\mathrm{FB}}$ and
$F_{\PH}$ as a function of the dimuon mass squared ($q^2$) based on an angular fit of the
decay \Sig in proton-proton collisions at
$\sqrt{s} = 8\TeV$. Charge-conjugate decay modes are implied throughout
this paper. The data, corresponding to an integrated luminosity of
20.5\fbinv~\cite{LUMI}, were collected by the CMS experiment at the
LHC in 2012. The angular
distribution of this decay has previously been studied by the
BABAR~\cite{Babar2006}, Belle~\cite{Belle2009}, CDF~\cite{CDF2012},
and LHCb~\cite{LHCb2012An,LHCb2014An} experiments, but no hints of BSM
have been seen.

\section{The CMS detector}

The central feature of the CMS detector is a superconducting solenoid
of 6\unit{m} internal diameter, providing a magnetic field of
3.8\unit{T}. Within the solenoid volume are a silicon pixel and a strip
tracker, a lead tungstate crystal electromagnetic calorimeter, and a
brass and scintillator hadron calorimeter, each composed of a barrel
and two endcap sections. Forward calorimeters extend the
pseudorapidity coverage provided by the barrel and endcap
detectors. Muons are detected in gas-ionization chambers embedded in
the steel flux-return yoke outside the solenoid. A detailed
description of the CMS detector, together with a definition of the
coordinate system used and the relevant kinematic variables, can be
found elsewhere~\cite{CMS}.

The events are selected online using a two-stage trigger
system~\cite{Khachatryan:2016bia}. The first level is composed of
custom hardware processors and uses information from the calorimeters
and muon detectors to select events at a rate of around 100\unit{kHz}
within a time interval of less than 4\mus. The second level, known as
the high-level trigger (HLT), consists of a farm of processors running
a version of the full event reconstruction software optimized for fast
processing, and reduces the event rate to around 1\unit{kHz} before
data storage.

\section{Event selection}
\label{sec:dataset}

The data for this analysis was recorded using a low-mass dimuon HLT
with a displaced vertex. The trigger requires a pair of opposite-sign
muons with a dimuon vertex displaced from the
interaction point by more than three times the calculated uncertainty.
The trigger also requires the dimuon candidate to have invariant mass
in the range 1.0--4.8\GeV and $\pt > 6.9\GeV$,
and for each muon to have $\pt>3.5\GeV$ and $\abs{\eta}<2.2$.

Monte Carlo (MC) simulated event samples are widely used in the
analysis. The number of simulated events for the signal sample \Sig
corresponds to more than 160 times that of the data. Other simulated
samples used in this analysis are \JPS, \PsP, and \PSX. In the last
decay mode, the muon pairs come from \JPsi or $\Pgy$ decay, and \cmsPX
denotes all other final-state particles. The MC samples are produced
using the \PYTHIA generator~\cite{PYTHIA6} version 6.424. Decays of
$\PBp$ and \JPsi or $\Pgy$ mesons are processed by the
\EVTGEN~\cite{EvtGEN} version 9.1 program (with the default matrix
element for the signal), in which final-state radiation is generated
using \PHOTOS~\cite{Golonka:2005pn}. Particles are traced through a
detailed model of the detector with \GEANTfour~\cite{GEANT4},
producing signals similar to the actual detector responses. Particles
coming from other proton-proton collisions in the same or nearby beam
crossings (pileup) are simulated according to the data-taking
conditions, but their effects on this analysis are small.

The selected events are reconstructed through the decay into the fully
charged final state of one charged hadron and a pair of oppositely
charged muons. Events from the control channels \JPS and \PsP have the
same final state as the signal process \Sig, and are extensively used
to validate the analysis and to evaluate the systematic
uncertainties. The muons are reconstructed using information from the
silicon tracker and muon detector systems~\cite{MUONPAS}. They must
satisfy the offline muon identification criteria that are optimized
for low-\pt muons~\cite{MUONID}. Dimuon candidates are formed from two
oppositely charged muons matching the HLT criteria that triggered the
event readout. To discriminate signal events from background,
additional selection criteria on kinematic variables are used. The
following selection criteria are determined through a maximization of
the expected signal significance using MC signal events and the
surviving data events in the final $\PBp$ meson invariant mass fitting
region, 5.1--5.6\GeV. The charged hadron track must have $\pt>1.3\GeV$
and the distance of closest approach in the transverse plane of the
charged hadron trajectory to the interaction point, divided by its
uncertainty, must be greater than 3.3. The $\PBp$ meson candidate is
formed by combining a dimuon candidate with the charged hadron track
assumed to be a kaon. The event kinematic information is updated by
fitting these three tracks to a common vertex.  The chi-squared
probability of the vertex fit for the $\PBp$ candidate is required to
be greater than 12\%.  To further reduce the background, the distance
in the transverse plane between the $\PBp$ vertex and the interaction
point must be larger than 10.6 times its uncertainty. The cosine of
the angle in the transverse plane between the $\PBp$ momentum and a
vector from the interaction point to the $\PBp$ meson vertex must be
greater than 0.9997.  After applying the selection criteria, less than
1\% of the selected events contain multiple $\PBp$ candidates. In
these events, only the candidate with the highest $\PBp$ decay vertex
fit probability is retained.

Events with a dimuon invariant mass ($q$) close to the \JPsi or $\Pgy$
resonance region are rejected to remove this contamination from the
control channels, as in Ref.~\cite{CMSKstarmumu2013}. The \JPsi and
$\Pgy$ resonance regions are defined as
$\smash[b]{m_{\JPsi}^{\mathrm{PDG}}-5\sigma_{q} < q <
  m_{\JPsi}+3\sigma_{q}}$ and
$\smash[b]{\abs{q-m_{\Pgy}^{\mathrm{PDG}}} < 5\sigma_{q}}$,
respectively, where $\sigma_{q}$ is the calculated uncertainty in $q$,
and the PDG superscript indicates the world-average mass
value~\cite{PDG} for each particle. We further suppress such events by
requiring,
$\smash[b]{\abs{(m-m_{\PBp}^{\mathrm{PDG}})-(q-m_{\JPsi}^{\mathrm{PDG}})}
  > 0.13\GeV}$ and
$\smash[b]{\abs{(m-m_{\PBp}^{\mathrm{PDG}})-(q-m_{\Pgy}^{\mathrm{PDG}})}
  > 0.06\GeV}$ in the $\PBp$ meson invariant mass region of
5.1--5.6\GeV, where $m$ is the $\PBp$ candidate invariant mass.  With
these requirements, the maximum contribution of events containing a
\JPsi or $\Pgy$ is less than 7\% in any $q^2$, and the kinematic
distributions of these events can be described together with those of
the combinatorial background.

\section{Angular analysis}
\label{sec:strategy}

The measurement of $A_{\mathrm{FB}}$ and $F_{\PH}$ is performed
through angular analysis in seven $q^2$ ranges from 1 to
22$\GeV^2$. The $q^2$ ranges used in this analysis are the same as in
previous measurements~\cite{Belle2009,CDF2012,LHCb2012An},
facilitating the comparison. The \JPsi and $\Pgy$ regions,
corresponding to $q^2$ ranges of 8.68--10.09 and 12.86--14.18$\GeV^2$,
respectively, are used as control
regions~\cite{CMSKstarmumu2013,CMSKstarmumu2016}. Additionally, we
define an inclusive low-$q^{2}$ range of 1.00--6.00$\GeV^2$ in order
to compare the results to SM calculations with the best-controlled
theoretical uncertainty, and a full inclusive $q^{2}$ range of
1.00--22.00$\GeV^2$, excluding the control regions. The analysis for
these two ranges is performed with the same procedure as for the other
ranges.

The decay rate for the process \Sig depends on
$\cos\theta_{\ell}$, where $\theta_{\ell}$ is the angle between the
directions of the $\Pgmm$ and $\PKp$ in the dilepton rest frame.
The differential decay width $\Gamma_{\ell}$ with respected to
$\cos\theta_{\ell}$ can be
parametrized~\cite{Theory:2007,ref2000,ref2001} in terms of the observables of
interest $A_{\mathrm{FB}}$ and $F_{\PH}$ as:
\begin{linenomath}
\begin{equation} \label{eq-AngAFBFH}
  \begin{split}
  \frac{1}{\Gamma_{\ell}}\frac{\rd\Gamma_{\ell}}{\rd\cos\theta_\ell} = \frac{3}{4}(1-F_{\PH})(1-\cos^{2}\theta_{\ell}) + \frac{1}{2}F_{\PH} + A_{\mathrm{FB}}\cos\theta_{\ell}.
  \end{split}
\end{equation}
\end{linenomath}
The requirement for the decay rate to remain positive over all
possible lepton angles constrains the parameter space to the region
$0 \leq F_{\PH} \leq 3$ and
$\abs{A_{\mathrm{FB}}} \leq \min(1, F_{\PH}/2)$. The angular
observables $A_{\mathrm{FB}}$ and $F_{\PH}$ are extracted from
a two-dimensional extended unbinned maximum-likelihood fit to the
angular distribution of the selected $\PBp$ meson candidates in
each $q^{2}$ range. The unnormalized probability density function
(pdf) used in the two-dimensional fit is:
\begin{linenomath}
  \ifthenelse{\boolean{cms@external}}
  {
    \begin{equation} \label{eq:PDF}
      \begin{split}
        \text{pdf}(m, \cos\theta_{\ell}) = Y_{\mathrm{S}}  S_{\mathrm{m}}(m)  S_{\mathrm{a}}(\cos\theta_{\ell}) \epsilon(\cos\theta_{\ell}) \\+ Y_{\mathrm{B}} B_{\mathrm{m}}(m) B_{\mathrm{a}}(\cos\theta_{\ell}),
      \end{split}
    \end{equation}
  }
  {
    \begin{equation} \label{eq:PDF}
      \text{pdf}(m, \cos\theta_{\ell}) = Y_{\mathrm{S}} S_{\mathrm{m}}(m)  S_{\mathrm{a}}(\cos\theta_{\ell})  \epsilon(\cos\theta_{\ell}) + Y_{\mathrm{B}}  B_{\mathrm{m}}(m)  B_{\mathrm{a}}(\cos\theta_{\ell}),
    \end{equation}
  }
\end{linenomath}
where the two contributions on the righthand side correspond to the
parametrization of the signal and background. The parameters $Y_{\mathrm{S}}$
and $Y_{\mathrm{B}}$ are the yields of signal and background events,
respectively. The functions $S_{\mathrm{m}}(m)$ and
$S_{\mathrm{a}}(\cos\theta_{\ell})$ describe the signal invariant mass
and angular distributions, while $B_{\mathrm{m}}(m)$ and
$B_{\mathrm{a}}(\cos\theta_{\ell})$ are similar functions describing
the background. The function $\epsilon(\cos\theta_{\ell})$ is the
signal efficiency as a function of $\cos\theta_{\ell}$.

The signal distribution $S_{\mathrm{m}}(m)$ is modeled as the sum of
two Gaussian functions with a common mean, and
$S_{\mathrm{a}}(\cos\theta_{\ell})$ is given in
Eq.~(\ref{eq-AngAFBFH}). The background distribution
$B_{\mathrm{m}}(m)$ is modeled as a single exponential function, while
$B_{\mathrm{a}}(\cos\theta_{\ell})$ is parametrized as the sum of a
Gaussian function and a third- or fourth-degree polynomial, depending
on the particular $q^2$ range.

Many of the parameters in the final fit are set to a given value with
a Gaussian constraint that reflects the input uncertainty of the
value. For the $S_{\mathrm{m}}(m)$ function, the mean is constrained
to the world-average \PBp mass~\cite{PDG} and the widths and relative
fraction of the two Gaussians are constrained to the values found from
fitting simulated events. The parameters of the
$B_{\mathrm{a}}(\cos\theta_{\ell})$ function are obtained by fitting
the events in the \PBp meson invariant mass sideband regions of
5.10--5.21 and 5.35--5.46\GeV. The free parameters of the fit are
$Y_{\mathrm{S}}$, $Y_{\mathrm{B}}$, $A_{\mathrm{FB}}$,
and $F_{\PH}$, as well as the exponential
decay parameter of $B_{\mathrm{m}}(m)$.

\begin{figure*}[!htb]
  \centering
  \includegraphics[width=0.329\textwidth]{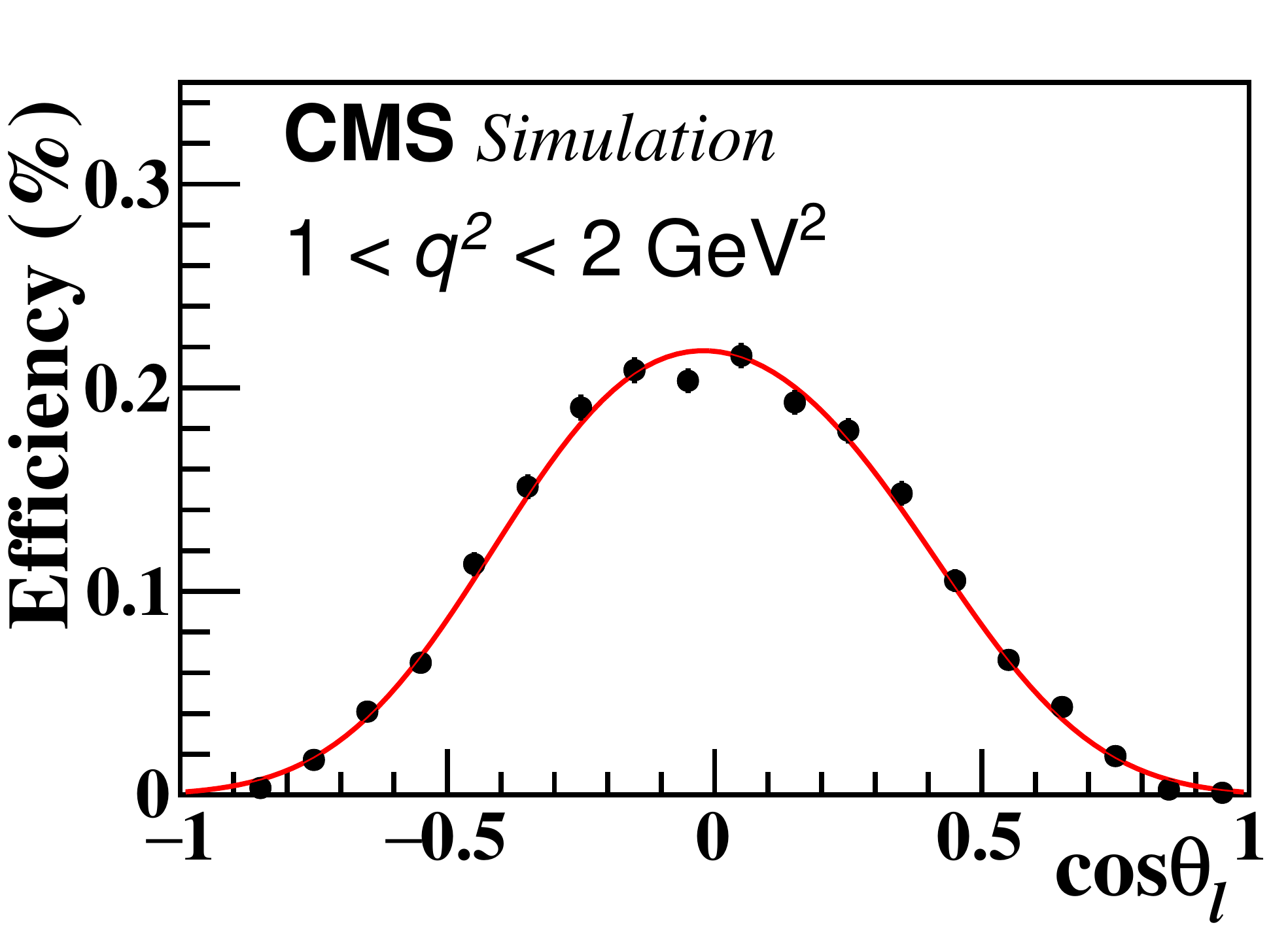}
  \includegraphics[width=0.329\textwidth]{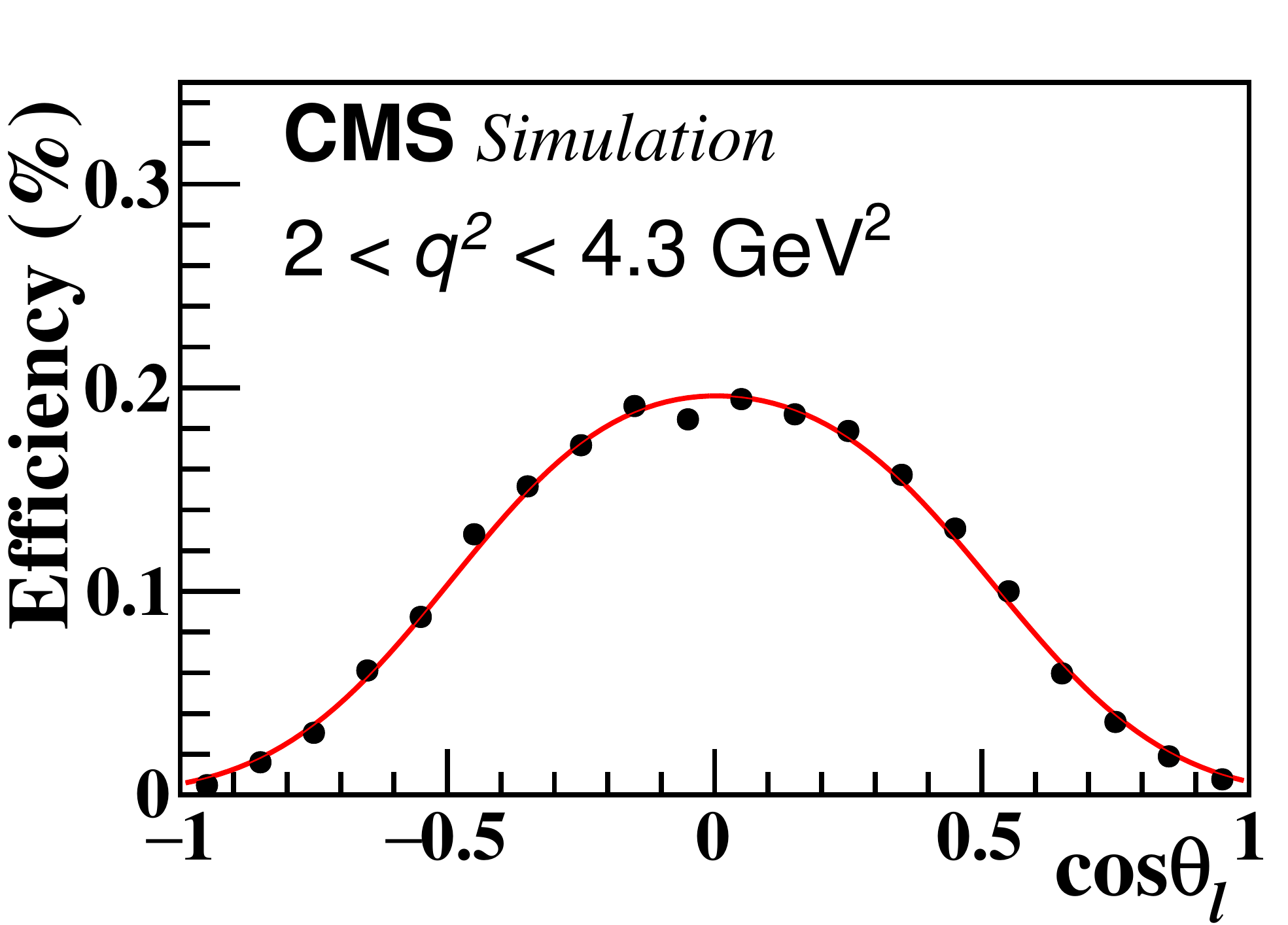}
  \includegraphics[width=0.329\textwidth]{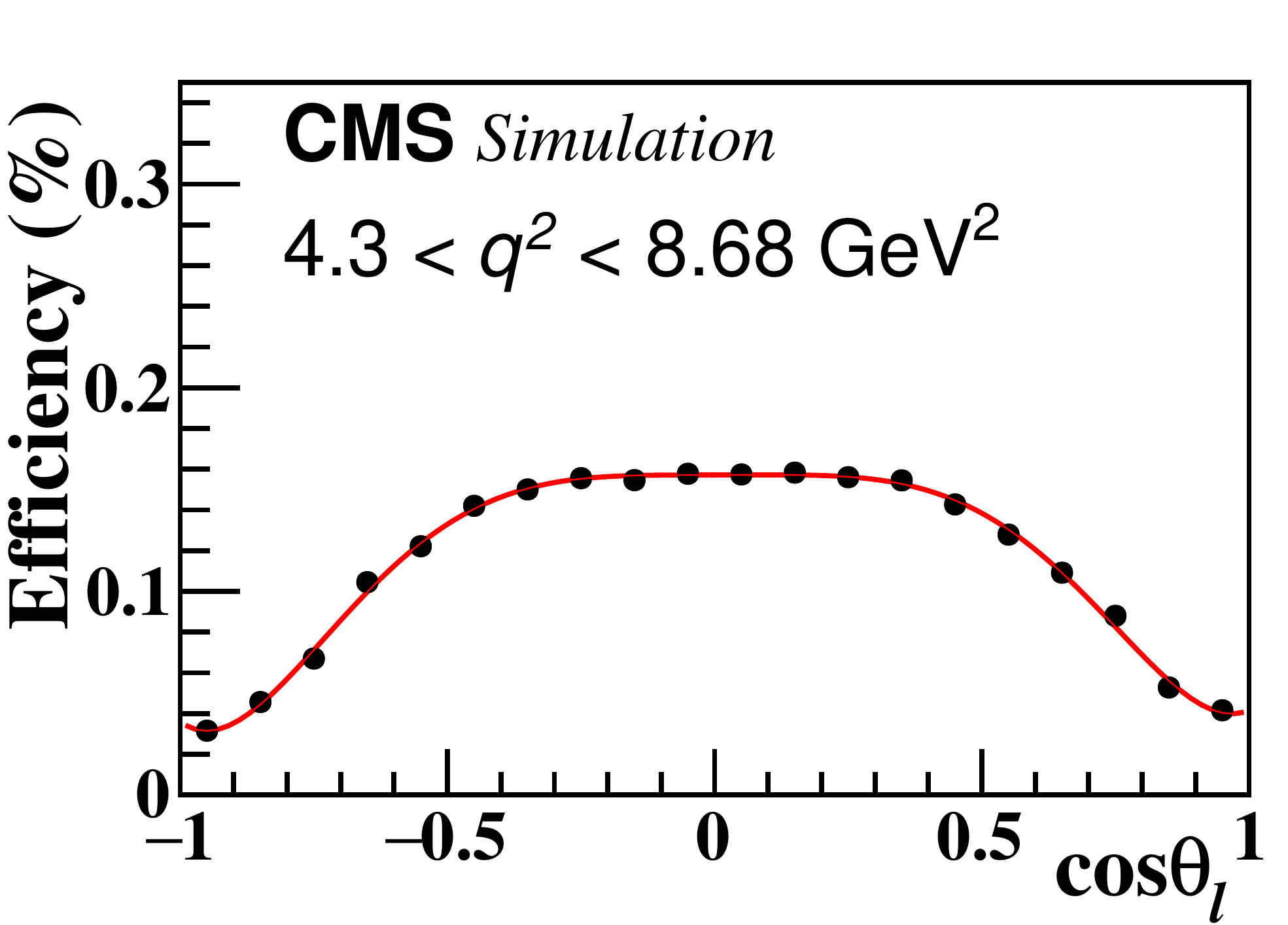}
  \includegraphics[width=0.329\textwidth]{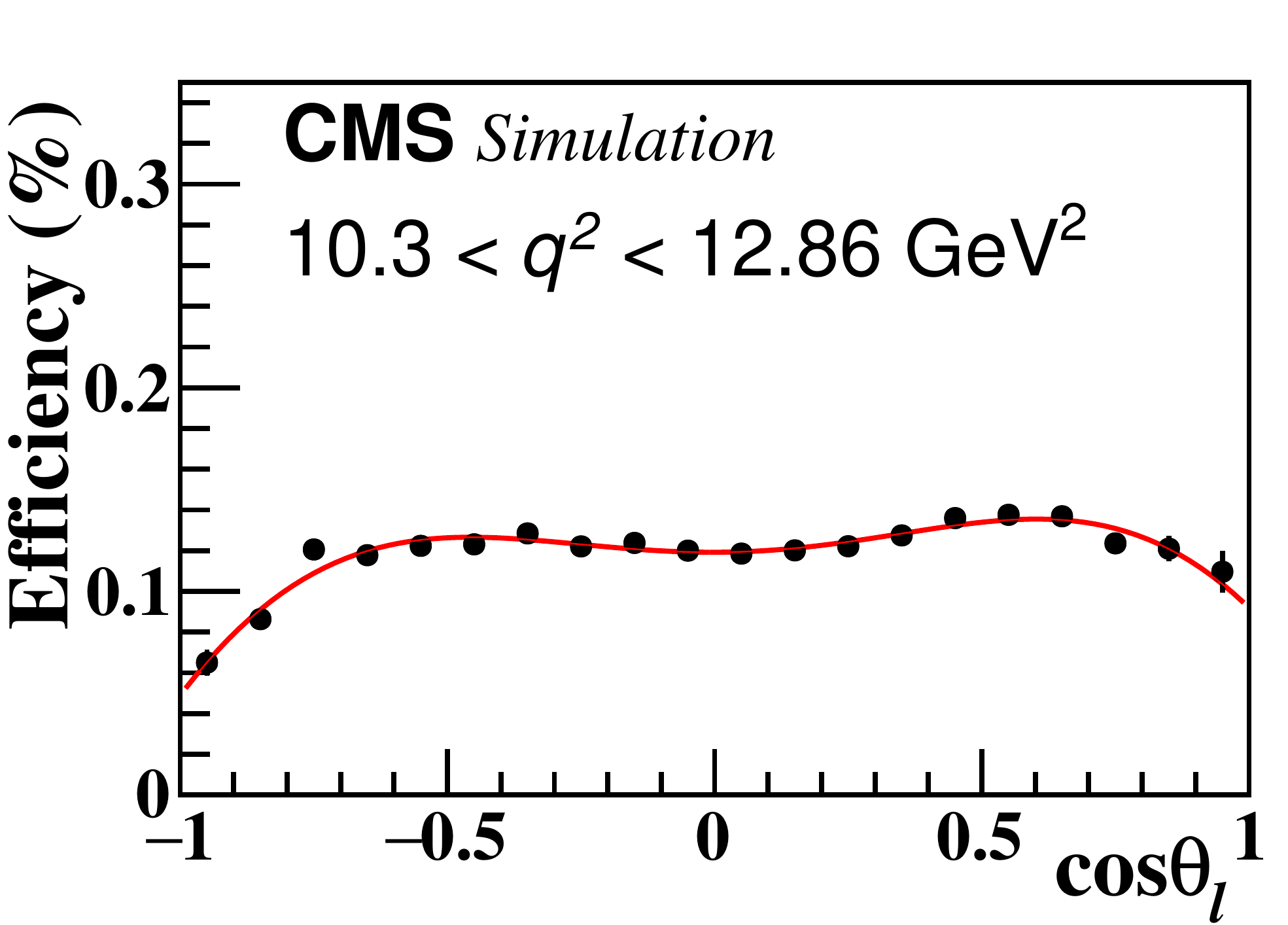}
  \includegraphics[width=0.329\textwidth]{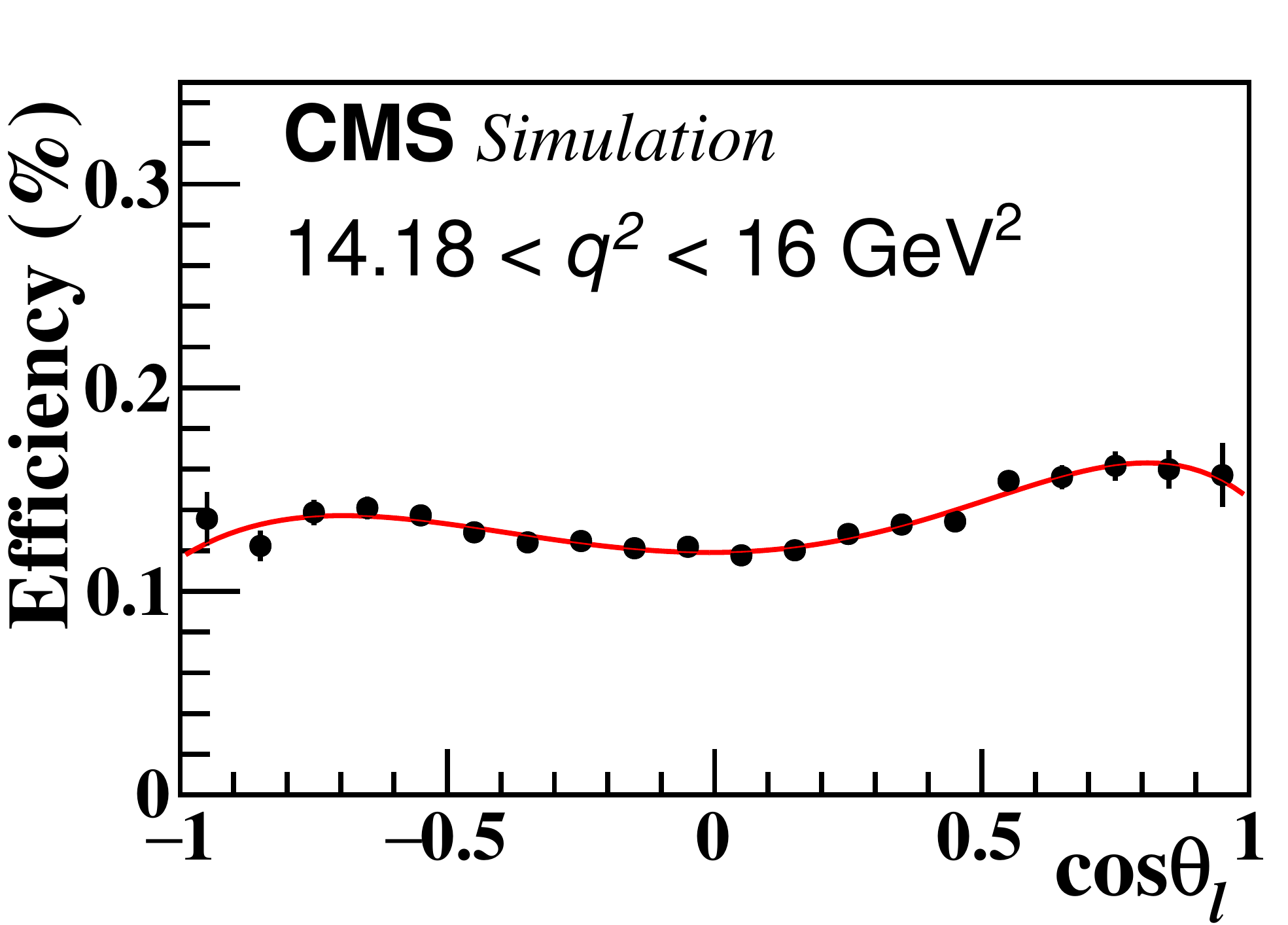}
  \includegraphics[width=0.329\textwidth]{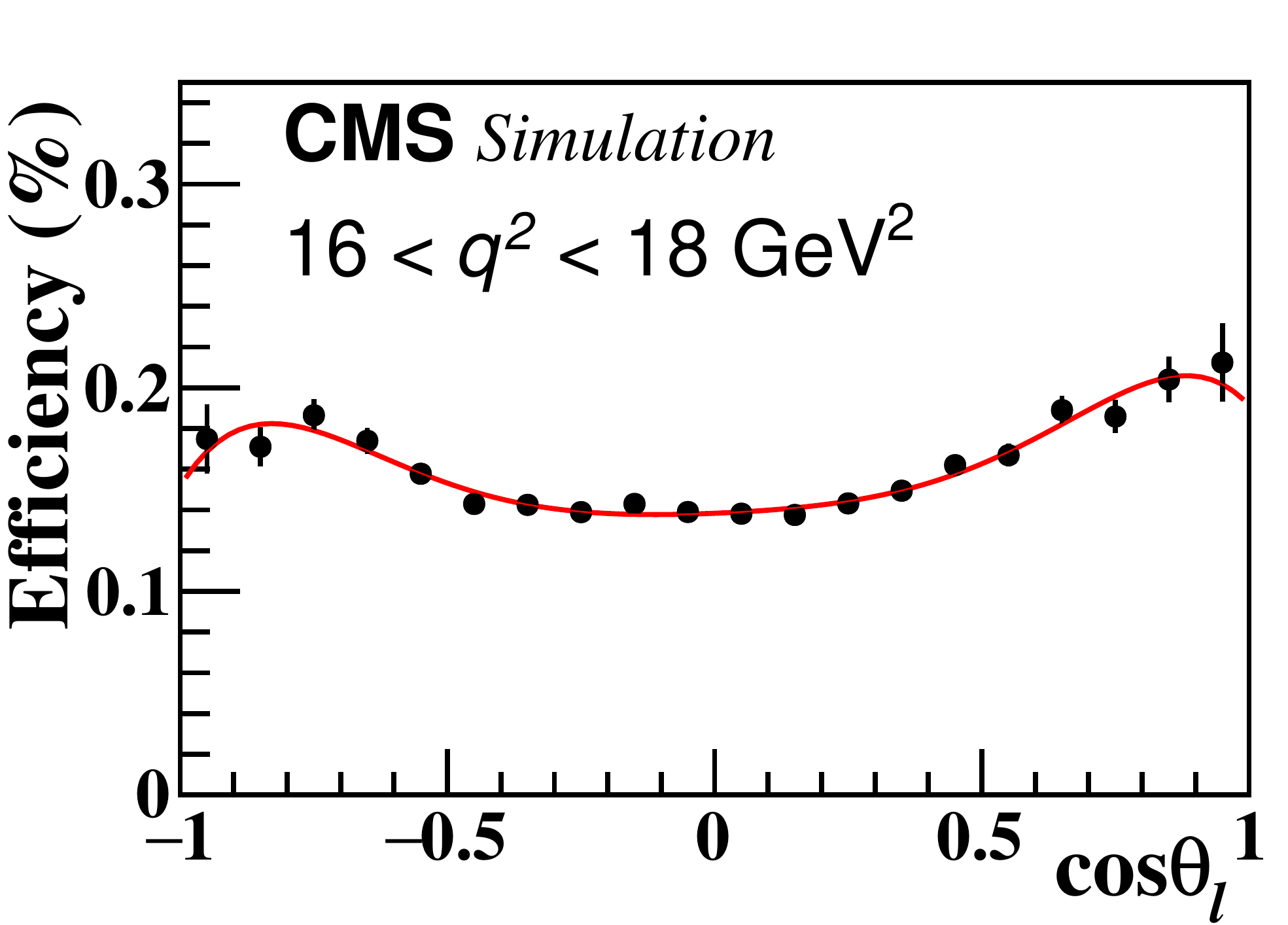}
  \includegraphics[width=0.329\textwidth]{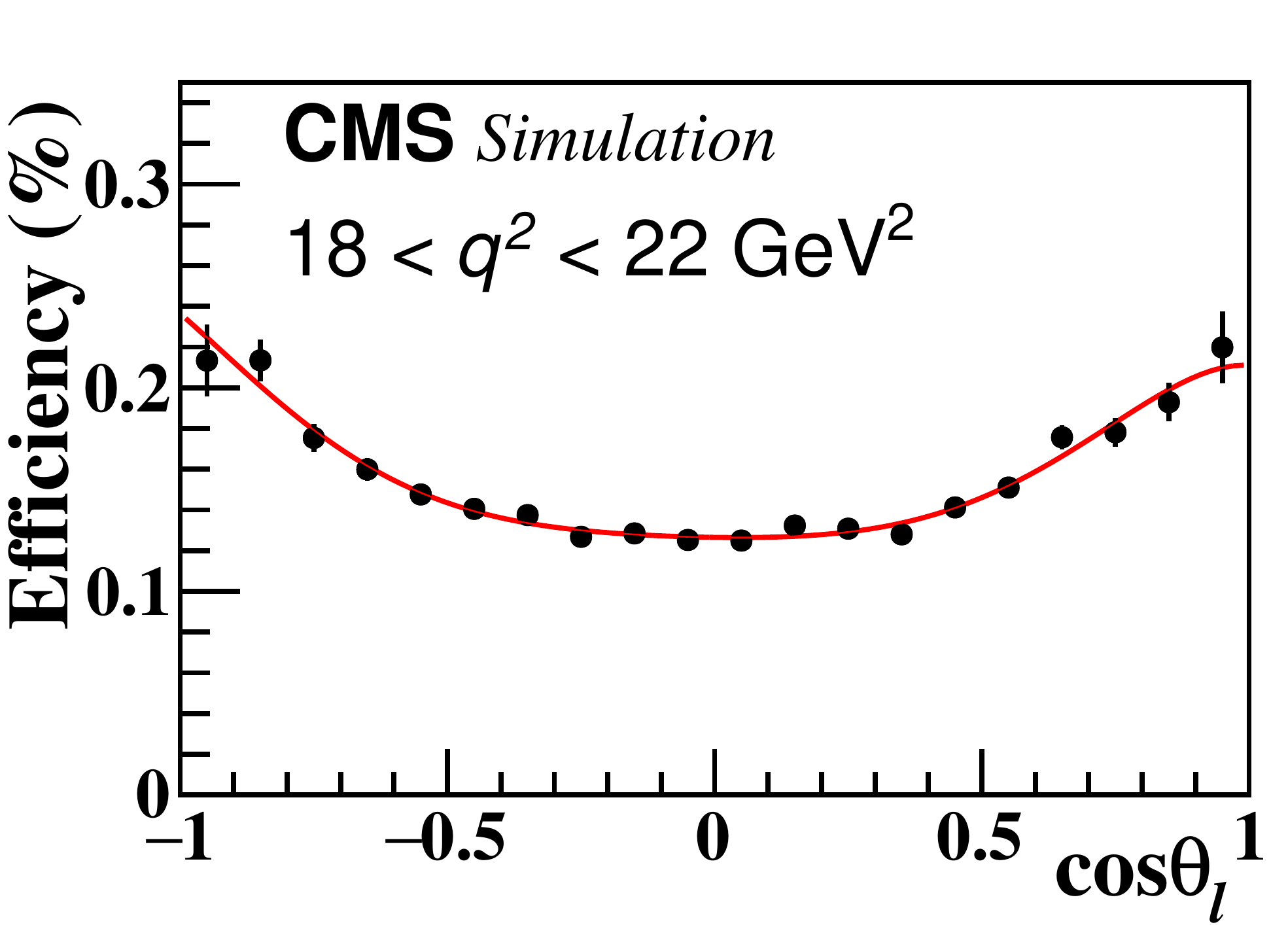}
  \includegraphics[width=0.329\textwidth]{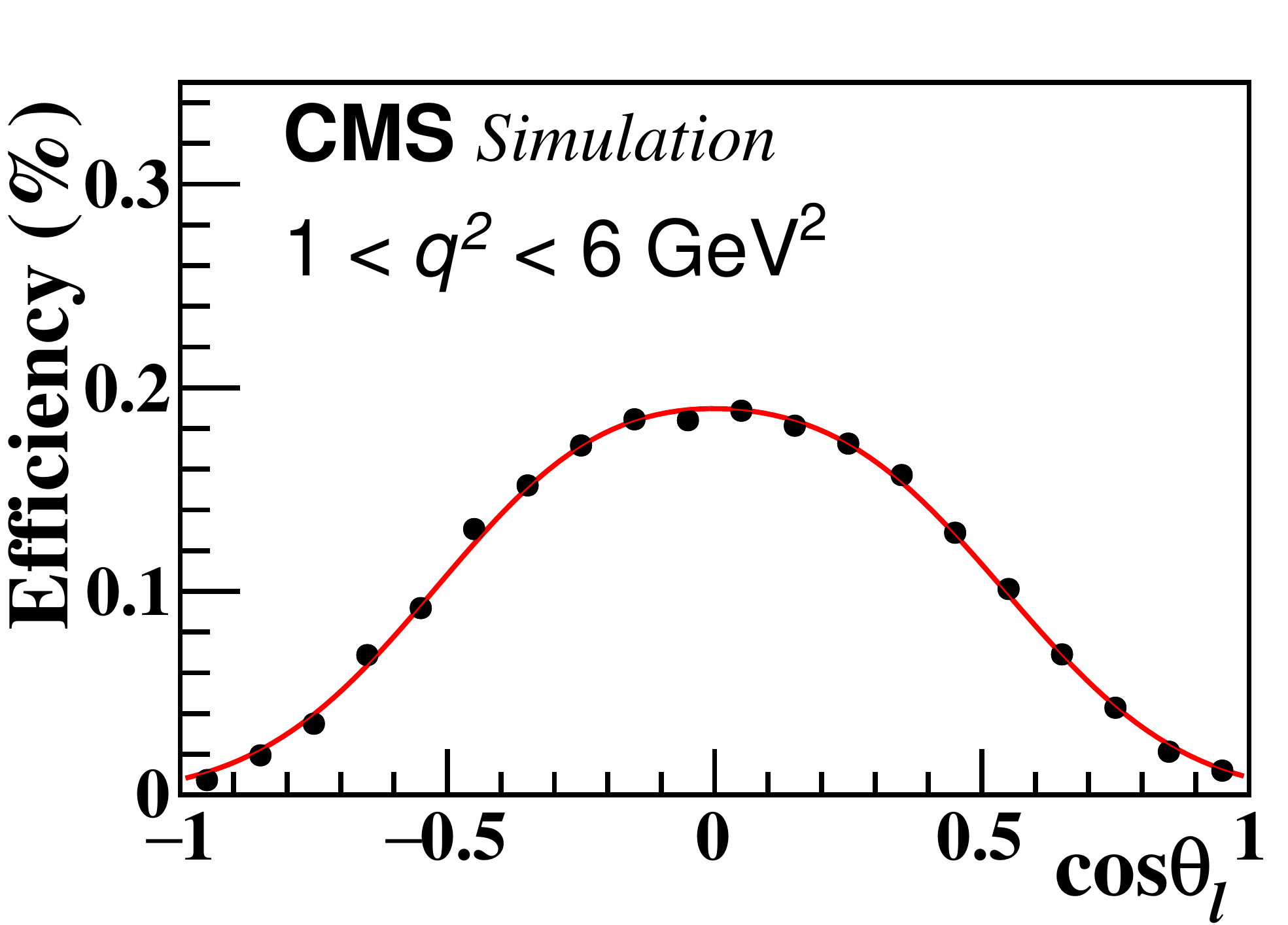}
  \includegraphics[width=0.329\textwidth]{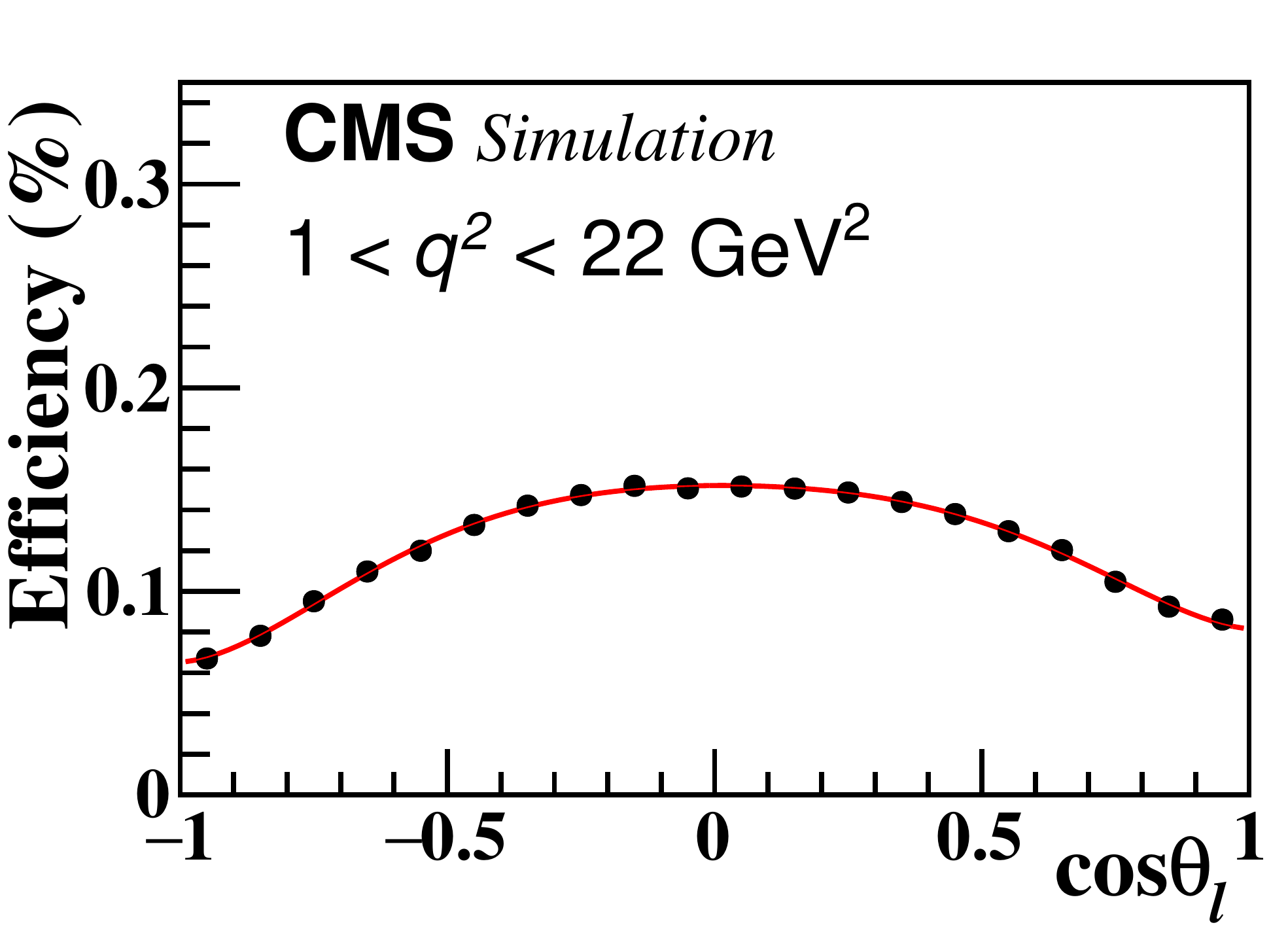}
  \caption{ The signal efficiency determined from simulated events as a function of
    $\cos\theta_{\ell}$ for the different $q^2$ ranges (points).
	The vertical bars indicate the statistical uncertainty.
	The curves show the sixth-order polynomial fits to the points.
  }
  \label{fig:eff_sig}
\end{figure*}

The signal efficiency $\epsilon(\cos\theta_{\ell})$ is factorized into
an acceptance $\epsilon_{\text{acc}}$ times a reconstruction
efficiency $\epsilon_{\text{reco}}$, which are both functions of
$\cos\theta_{\ell}$. The acceptance is obtained from generated
events, before the particle propagation with \GEANTfour, and is
calculated as the fraction of MC simulated signal events passing the
muon requirement of $\pt > 3.5\GeV$ and $\abs{\eta}<2.2$ relative to all
generated events. It varies from 2 to 4\% depending on $q^2$.
The reconstruction efficiency is obtained from the ratio of
the number of reconstructed MC events passing the final event
selection to the number of events passing the single-muon selection at
the generator level. It varies from 4 to 7\% depending on $q^2$.
The signal efficiency $\epsilon(\cos\theta_{\ell})$
is parametrized and fit with a sixth-order polynomial, as shown in
Fig.~\ref{fig:eff_sig} for the nine different signal $q^2$ ranges
used in this analysis.

The angular distributions of data and simulation from the two control
channels are compared and the good agreement between them provides a
validation of the efficiency description. We also check that the ratio
of the branching fractions of the two control channels is consistent
with the world-average value~\cite{PDG} within their uncertainties.
The MC simulation samples are used to validate the fitting procedure
in each $q^2$ range. The results of fitting the signal MC sample at
the generator level and the standard signal simulation are consistent
with each other. The large MC signal sample is divided into 20
subsamples and fits of these subsamples reveal no additional bias. In
addition, we generate 200 pseudo-experiments of 100 times the size of
data, using the pdf in Eq.~(\ref{eq:PDF}), with parameters from
fitting the data. The differences between the fitted values from these
samples and the input parameters from data follow Gaussian
distributions with the means consistent with zero and the widths
smaller than the variations among the signal MC subsample fits in the
same $q^2$ range.

\begin{figure*}[!htb]
    \centering
	\includegraphics[width=0.329\textwidth]{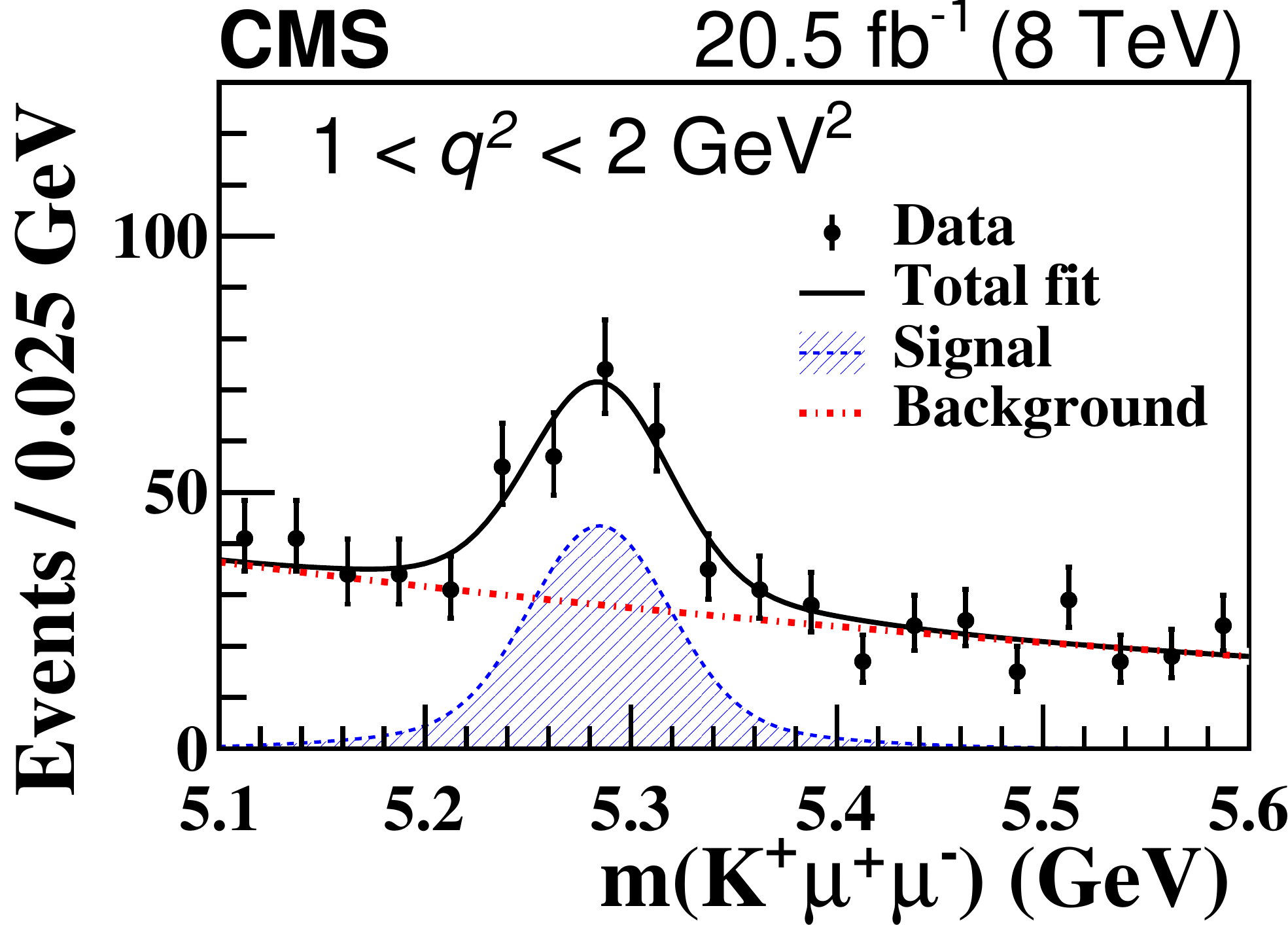}
    \includegraphics[width=0.329\textwidth]{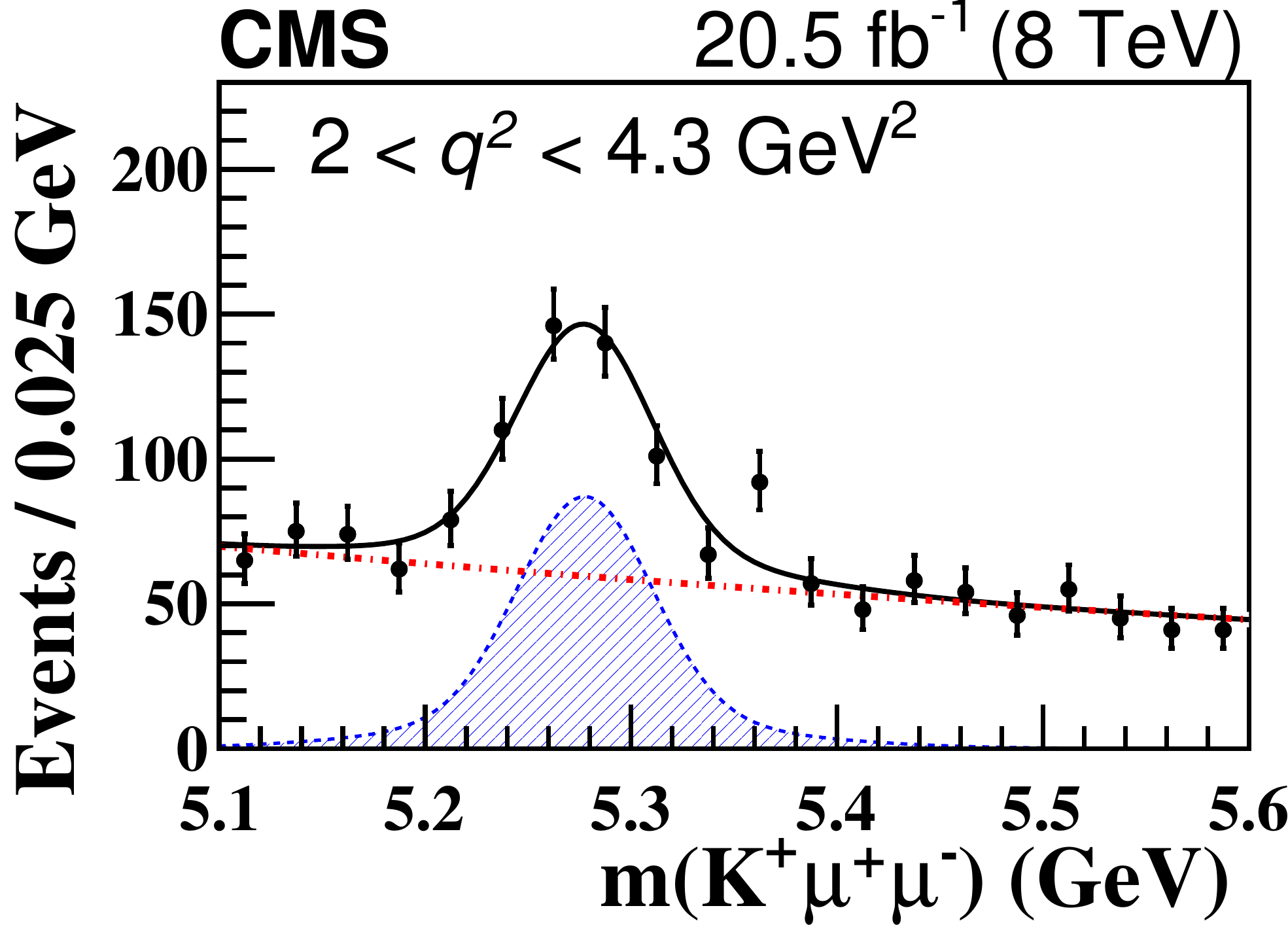}
    \includegraphics[width=0.329\textwidth]{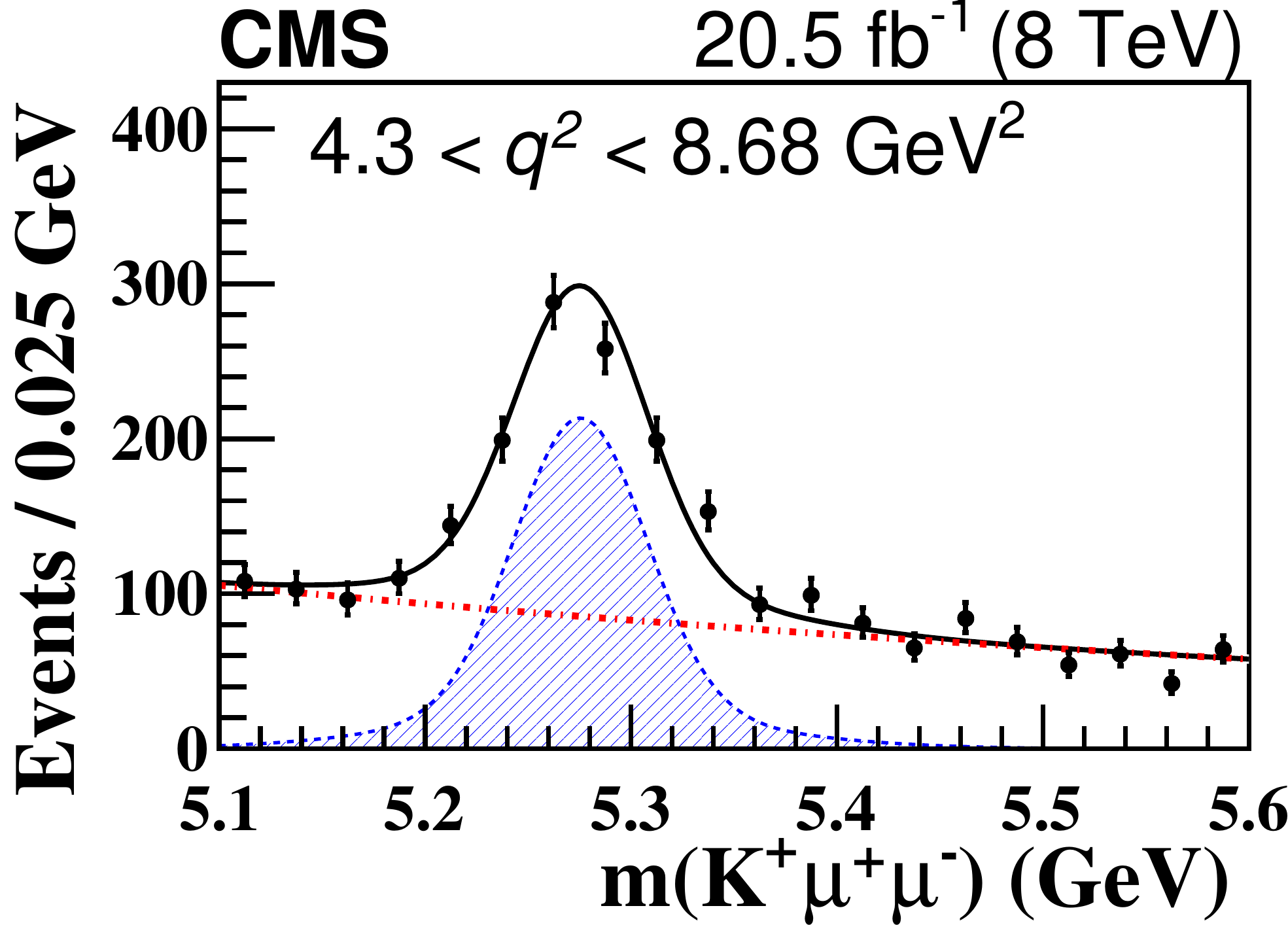}
    \includegraphics[width=0.329\textwidth]{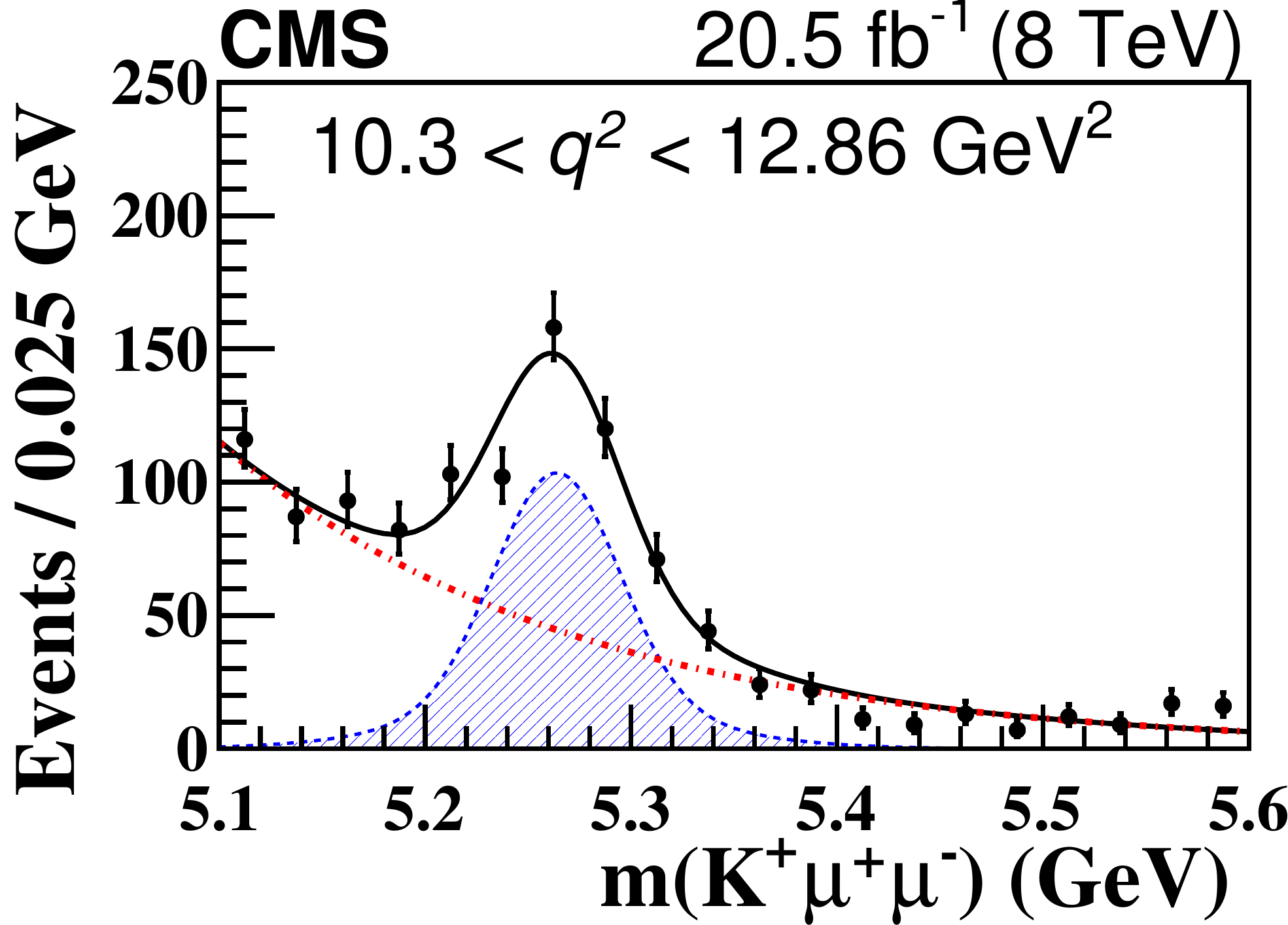}
    \includegraphics[width=0.329\textwidth]{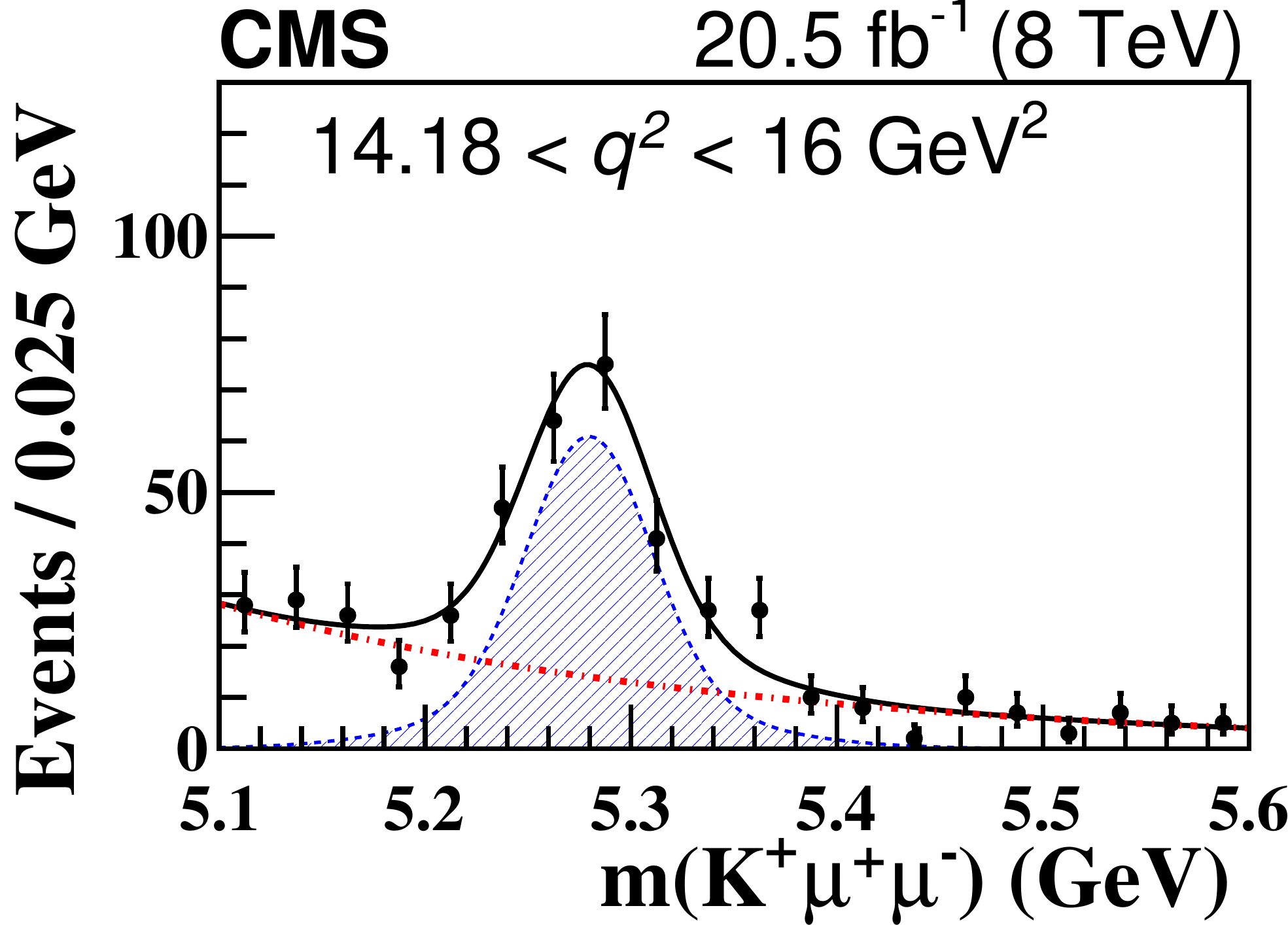}
    \includegraphics[width=0.329\textwidth]{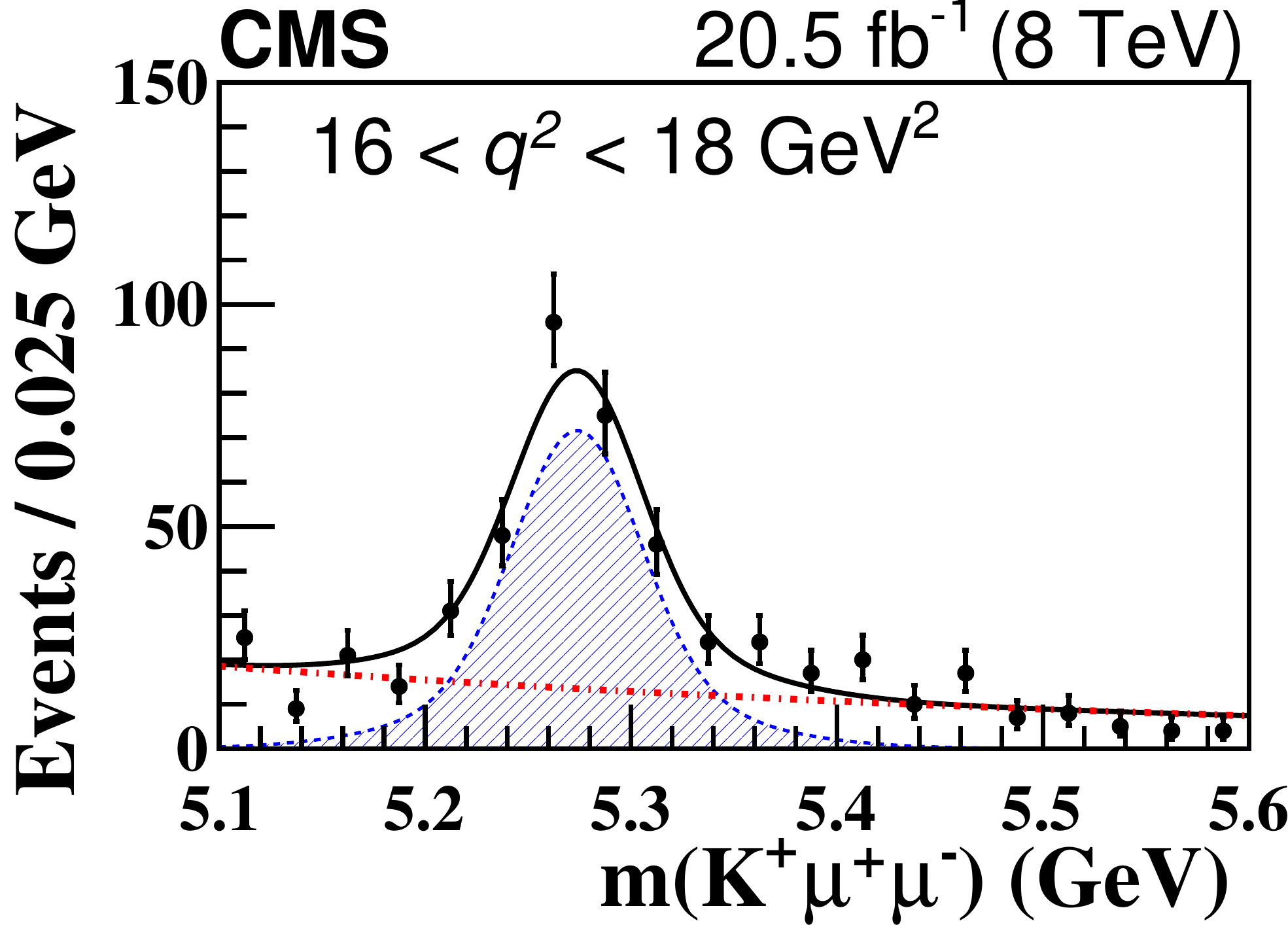}
    \includegraphics[width=0.329\textwidth]{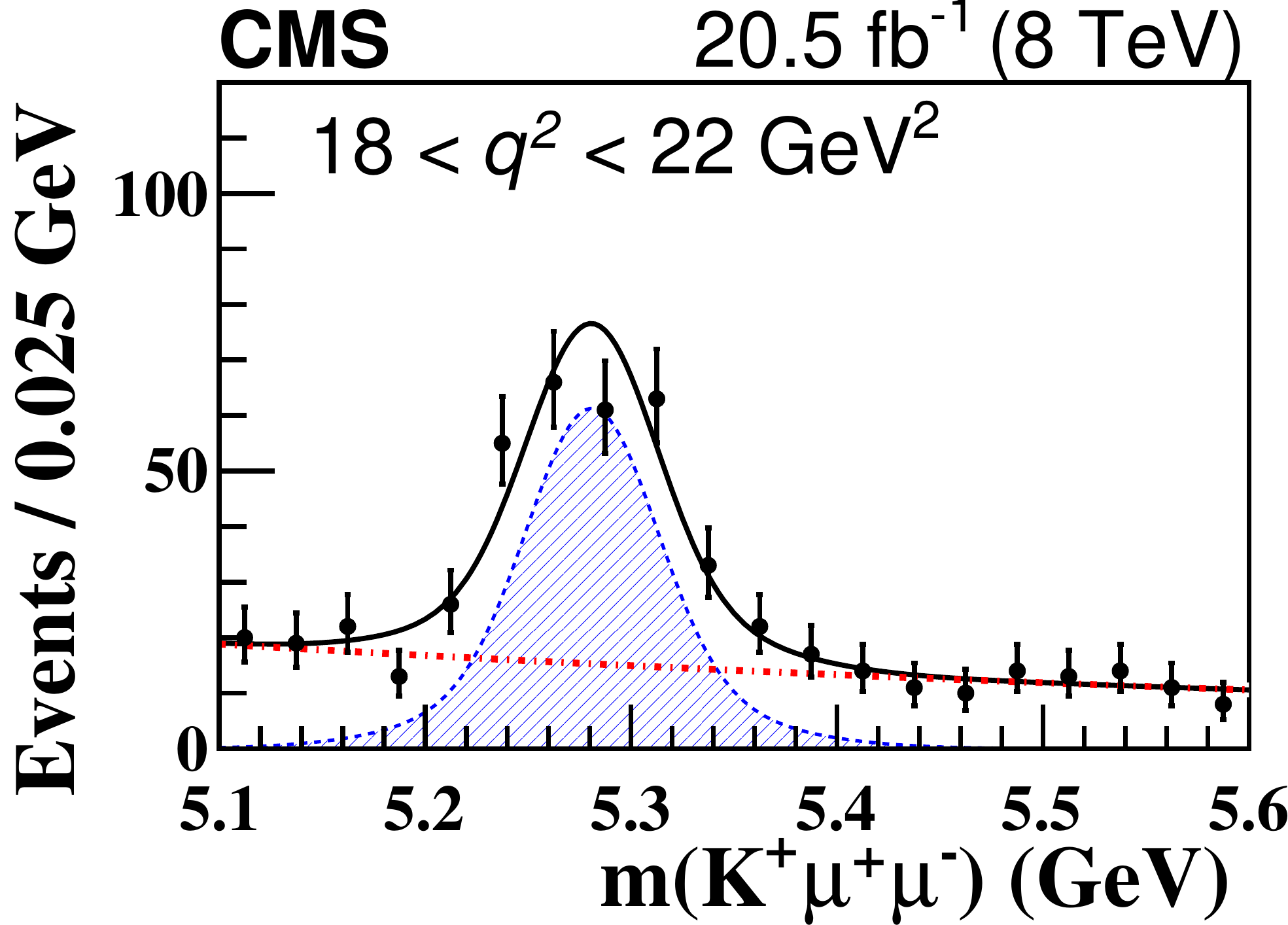}
    \includegraphics[width=0.329\textwidth]{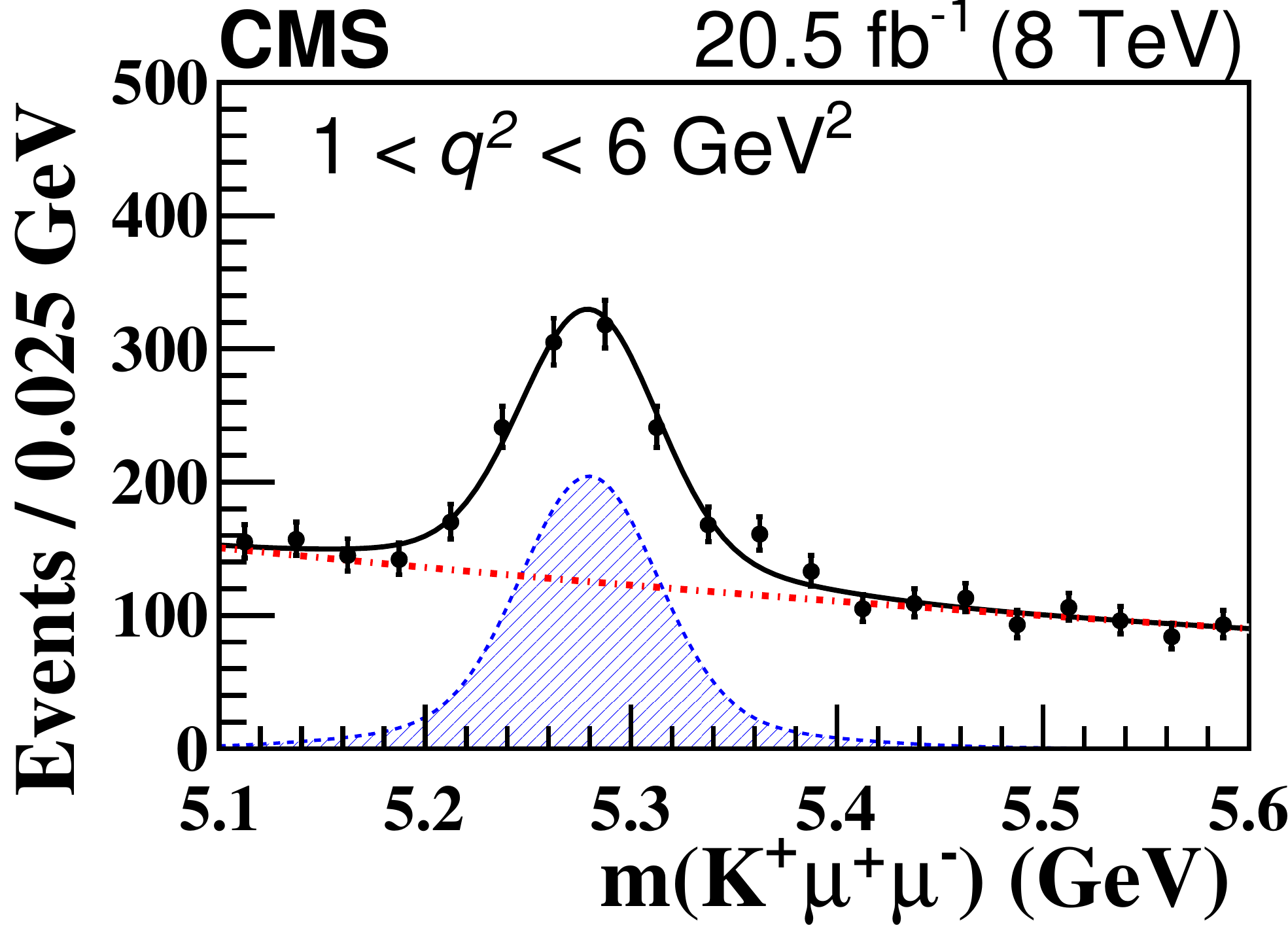}
    \includegraphics[width=0.329\textwidth]{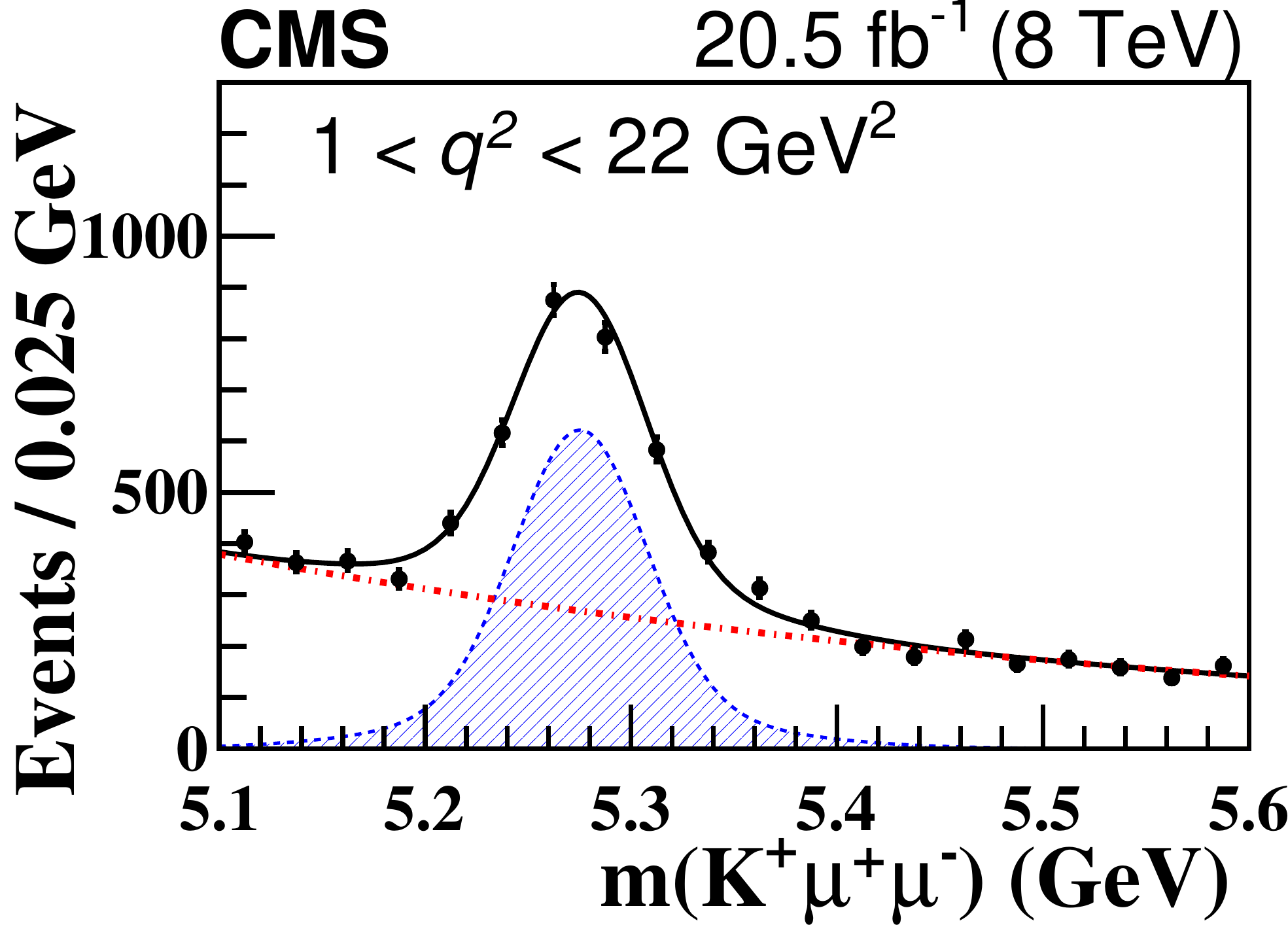}
    \caption{Projections of the $\PKp\Pgmp\Pgmm$ invariant
      mass distributions for each $q^2$ range from the two-dimensional
      fit of data. The solid lines show the total fit, the shaded area
      the signal contribution, and the dash-dotted lines the
      background. The vertical bars on the points represent the
      statistical uncertainty in data. }
	 \label{fig:FHAFBdatabmass}
\end{figure*}

\begin{figure*}[!htb]
    \centering
	\includegraphics[width=0.329\textwidth]{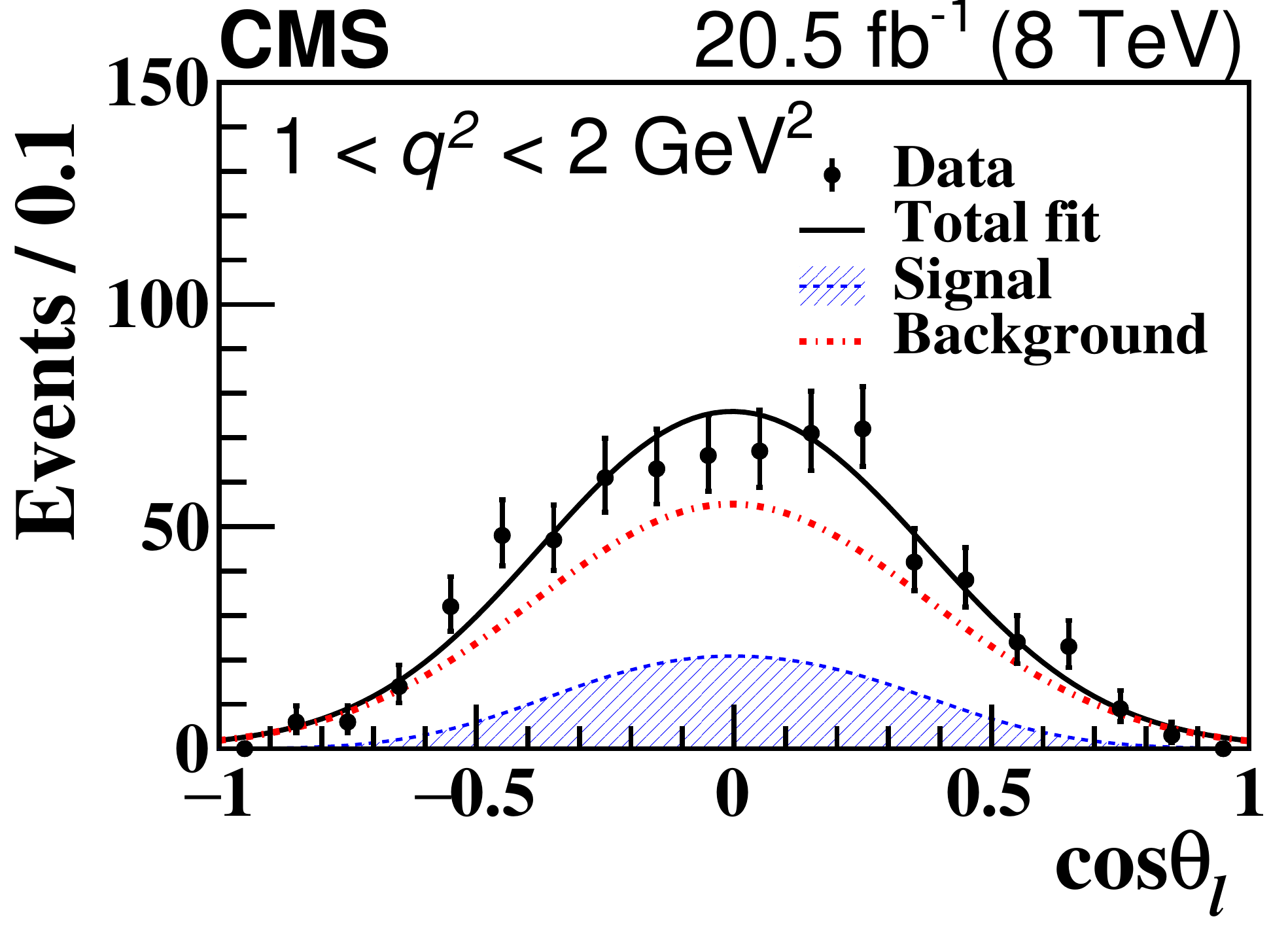}
    \includegraphics[width=0.329\textwidth]{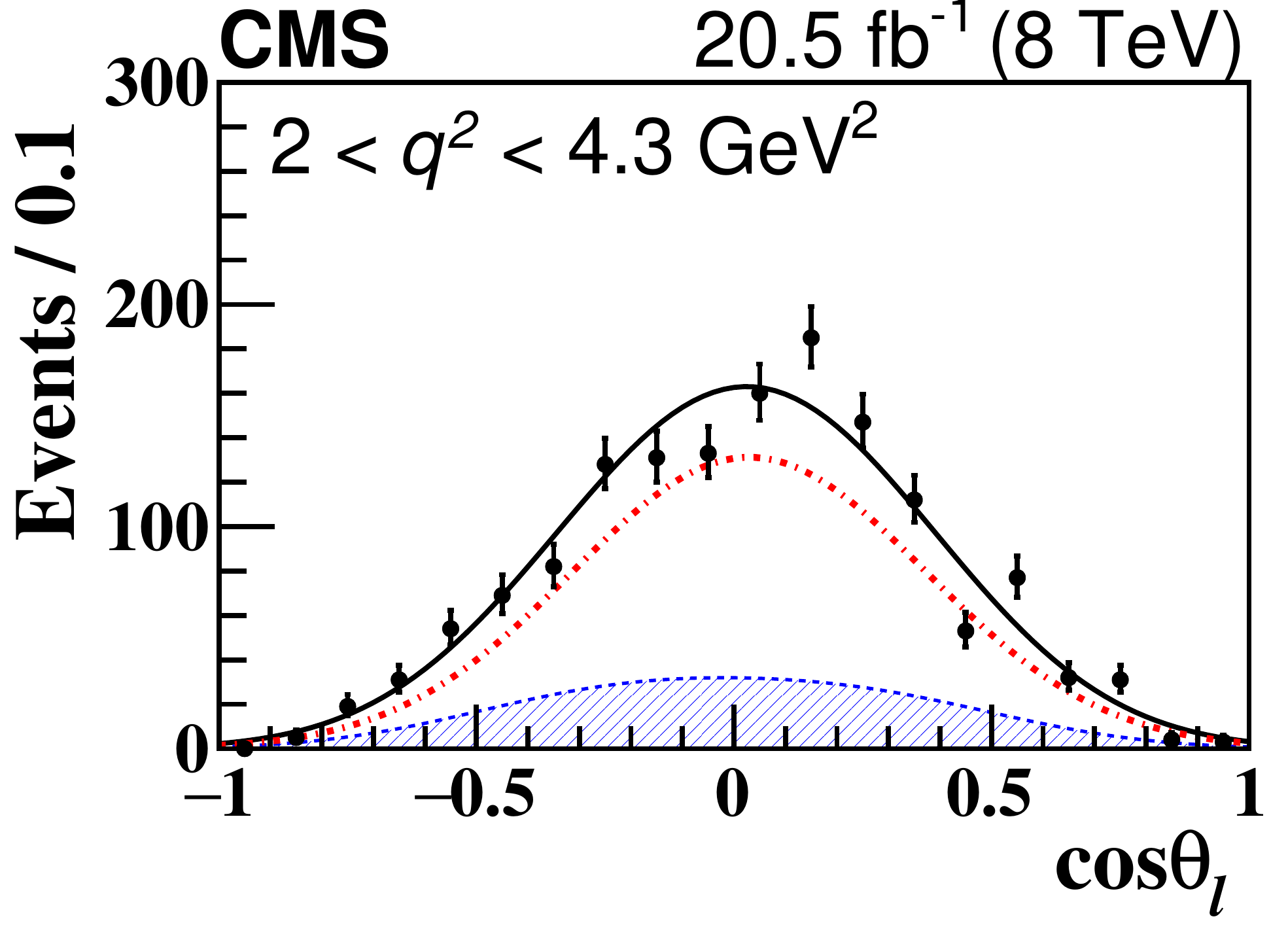}
    \includegraphics[width=0.329\textwidth]{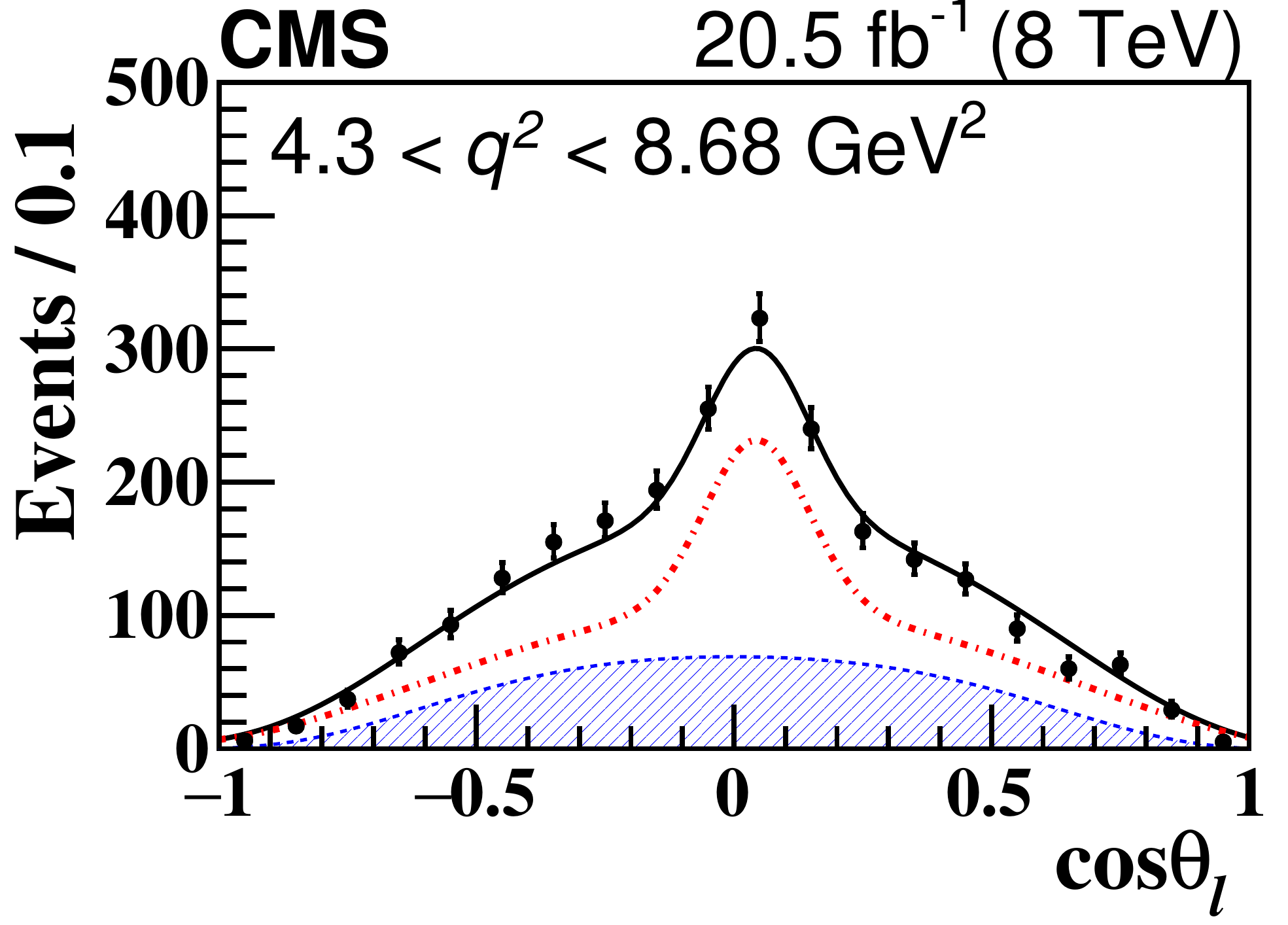}
    \includegraphics[width=0.329\textwidth]{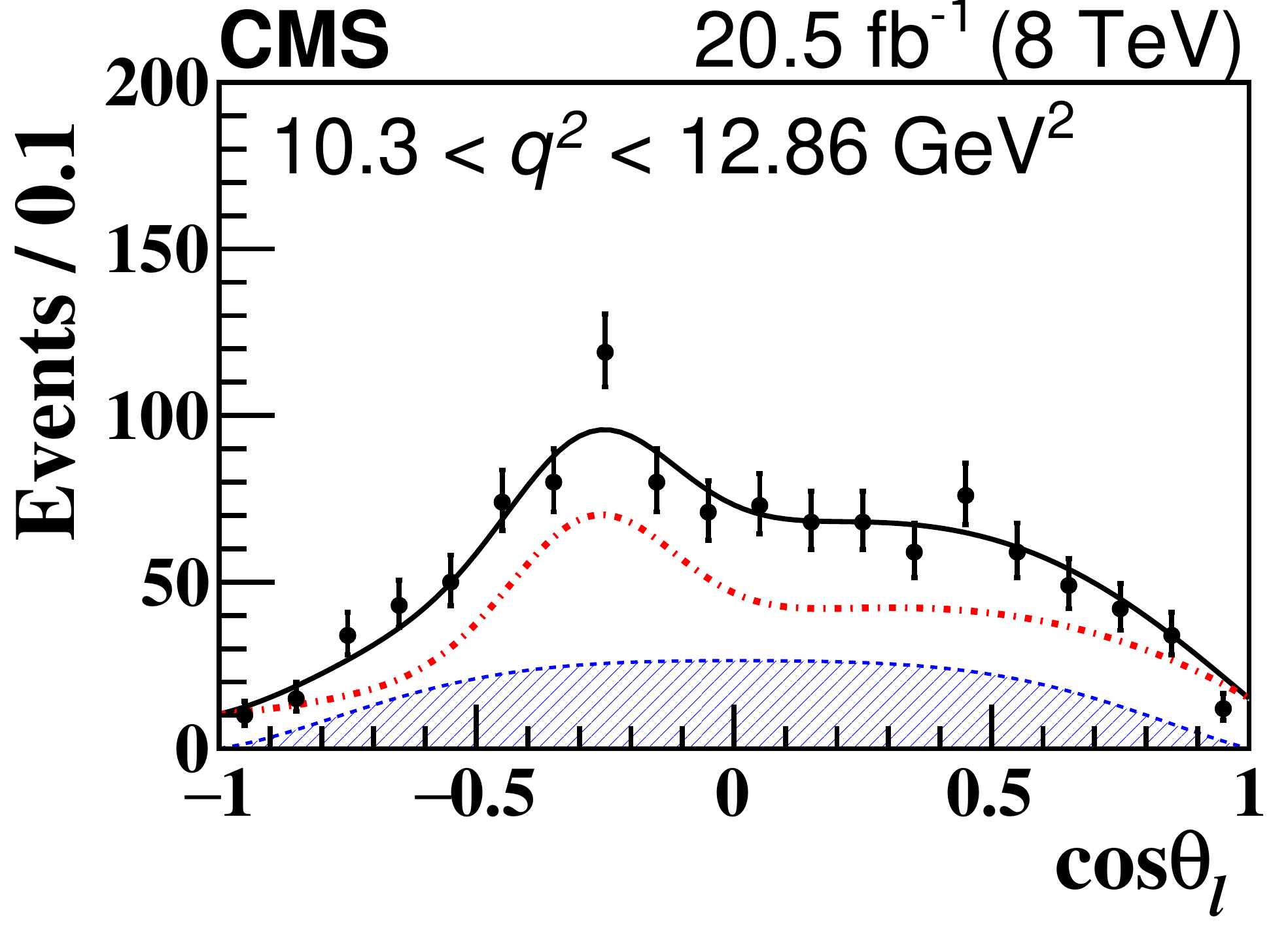}
    \includegraphics[width=0.329\textwidth]{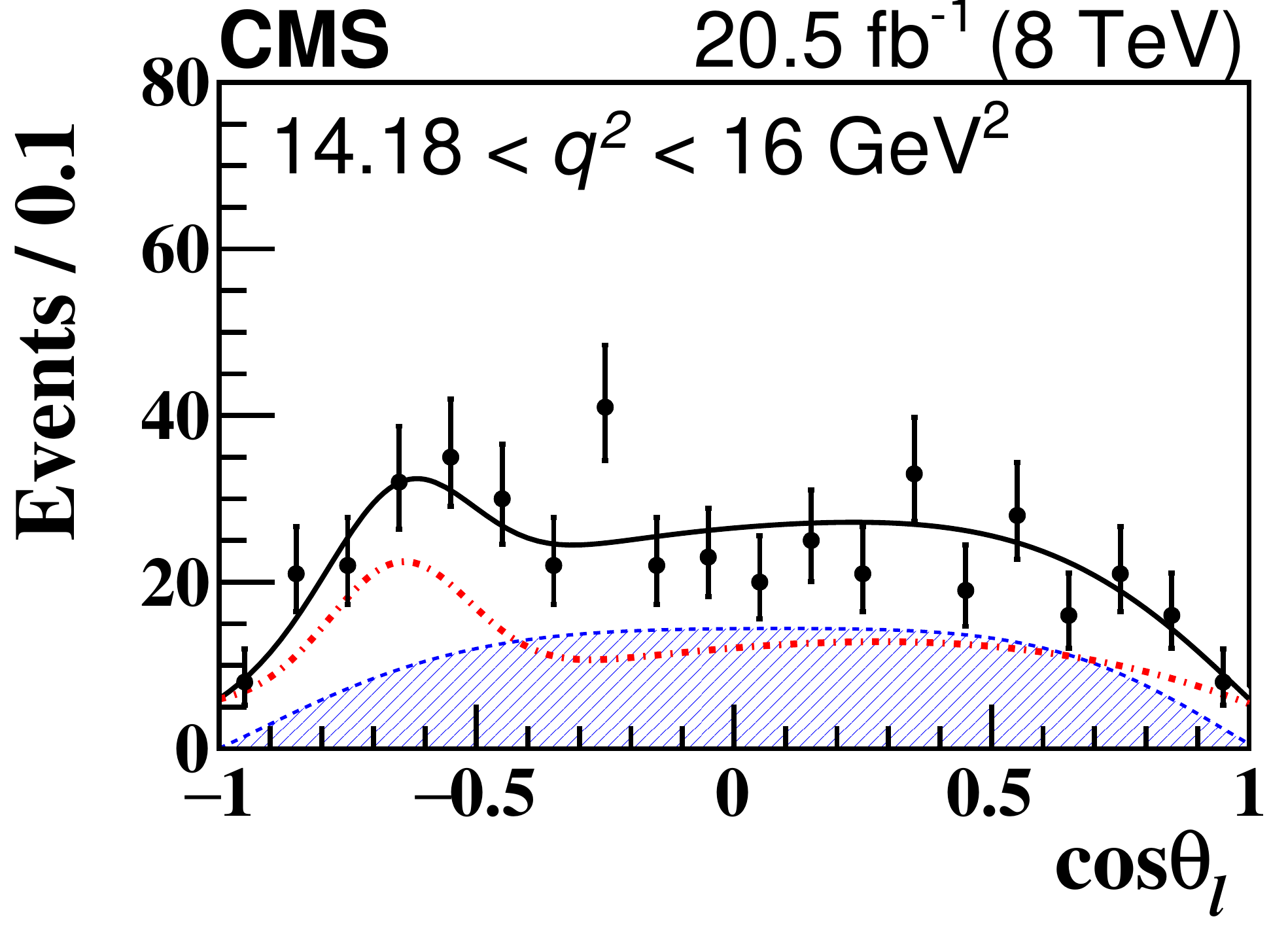}
    \includegraphics[width=0.329\textwidth]{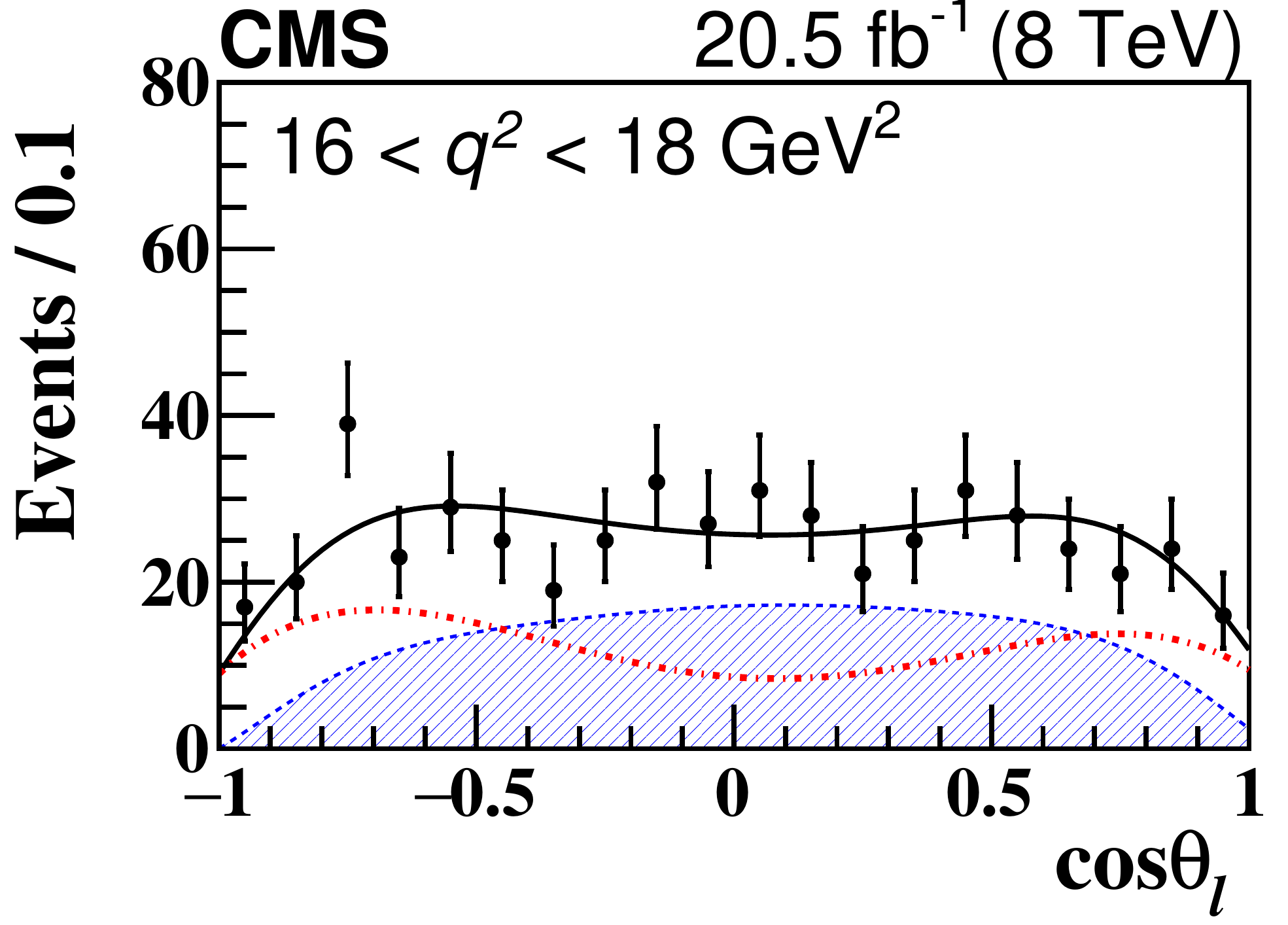}
    \includegraphics[width=0.329\textwidth]{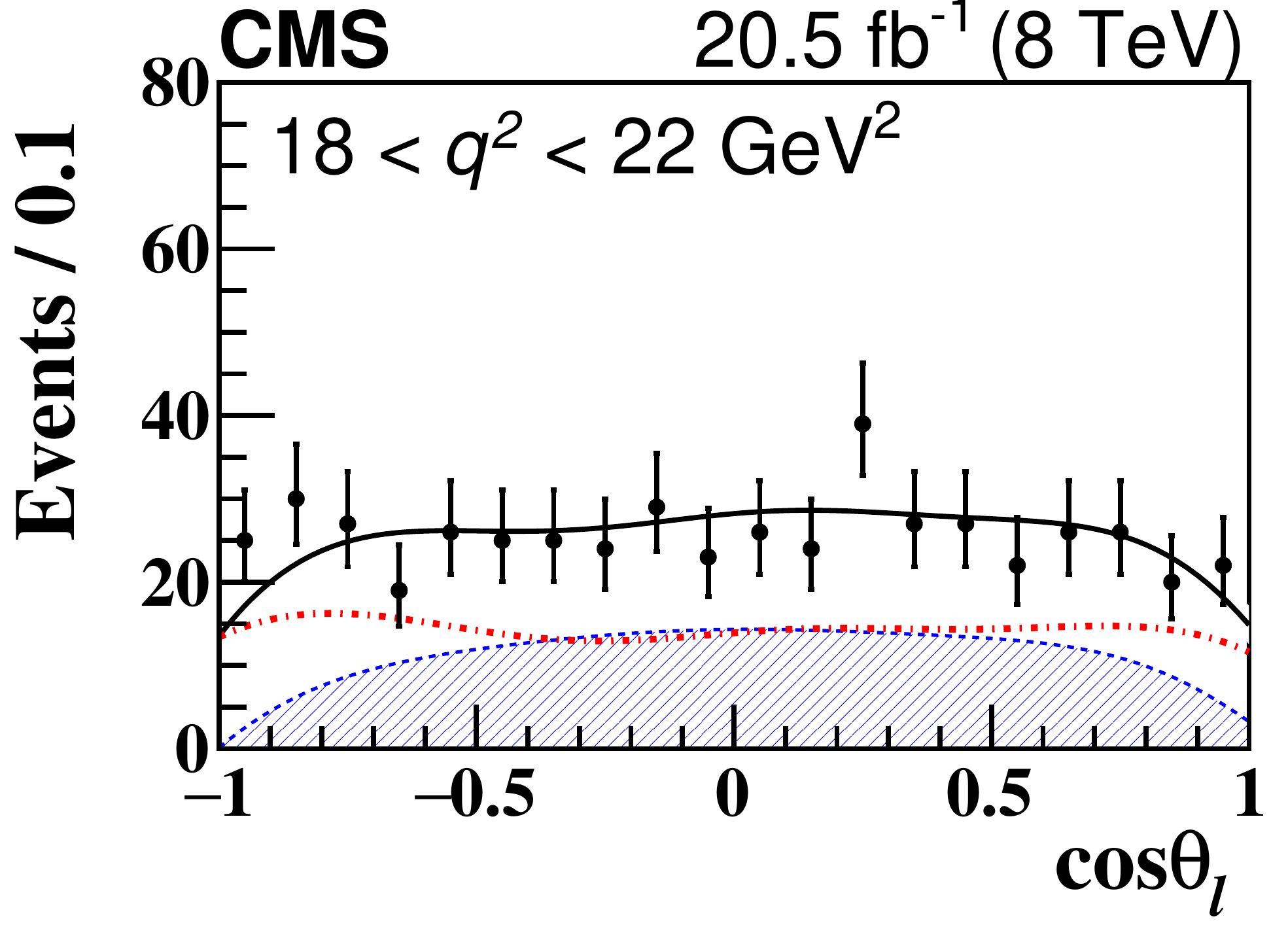}
    \includegraphics[width=0.329\textwidth]{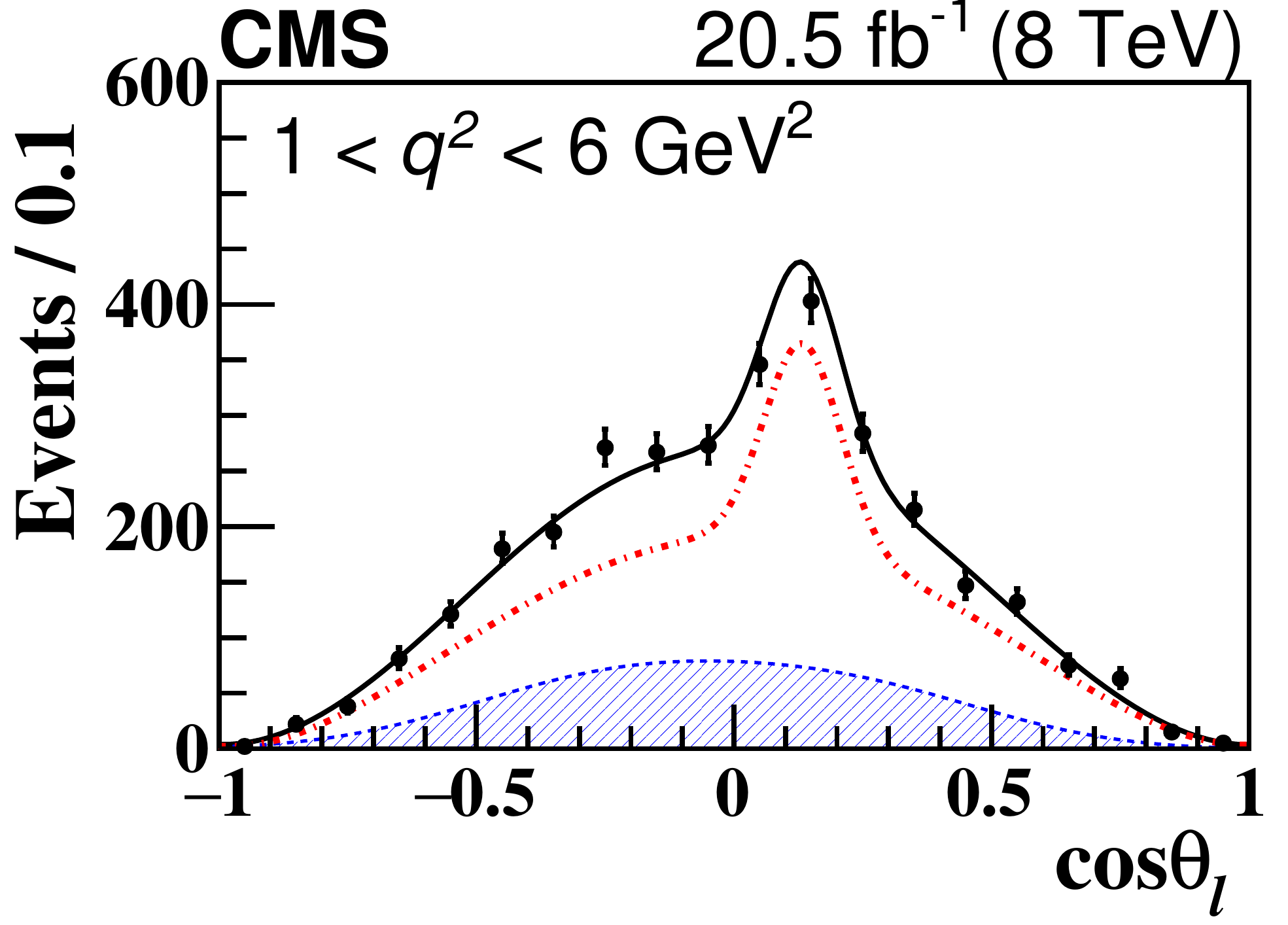}
    \includegraphics[width=0.329\textwidth]{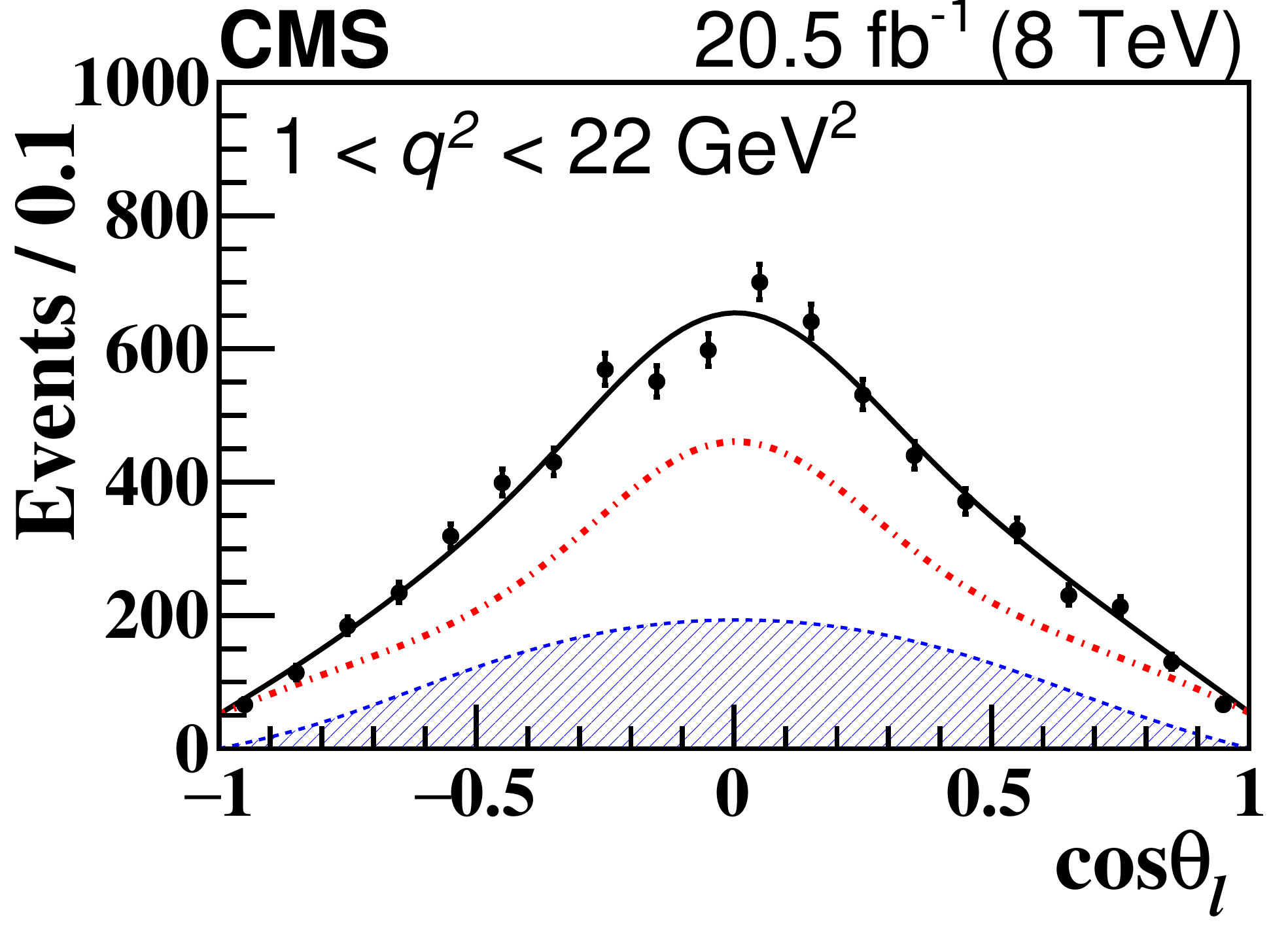}
    \caption{Projections of the $\cos\theta_{\ell}$ distributions for
      each $q^2$ range from the two-dimensional fit of data. The solid
      lines show the total fit, the shaded area the signal
      contribution, and the dash-dotted lines the background. The
      vertical bars on the points represent the statistical
      uncertainty in data. }
	 \label{fig:FHAFBdatacosl}
\end{figure*}

The final fit is performed over the full $\PBp$ meson invariant
mass range and results in $2286\pm73$ signal events with $q^2$ from 1 to
22$\GeV^2$. Figures~\ref{fig:FHAFBdatabmass} and
~\ref{fig:FHAFBdatacosl} show the $\PKp\Pgmp\Pgmm$ invariant
mass and the $\cos\theta_{\ell}$ projections, respectively, for each $q^2$
range from the two-dimensional fit to the data.

\section{Systematic uncertainties}
\label{sec:systematic}

Several sources of systematic uncertainty in the measured values of $A_{\mathrm{FB}}$ and
$F_{\PH}$ are considered, as summarized in Table~\ref{tab:sysAFBFH}.
Varying the parameter values of $S_{\mathrm{m}}(m)$ used to fit
the signal invariant mass distribution within their
uncertainties results in a negligible change in the measured values of
$A_{\mathrm{FB}}$ and $F_{\PH}$.

\begin{table}[!htb]
  \centering
  \topcaption{Absolute values of the uncertainty
    contributions in the measurements of $A_{\mathrm{FB}}$ and
    $F_{\PH}$. For each item, the range indicates the variation
    of the uncertainty in the signal $q^2$ ranges.}
  \begin{scotch}{lcc}
    Systematic uncertainty & $A_{\mathrm{FB}}~(\ten{-2})$ & $F_{\PH}~(\ten{-2})$ \\
    \hline
    Finite size of MC samples         & 0.4--1.8 & 0.9--5.0 \\
    Efficiency description            & 0.1--1.5 & 0.1--7.8 \\
    Simulation mismodeling            & 0.1--2.8 & 0.1--1.4 \\
    Background parametrization model  & 0.1--1.0 & 0.1--5.1 \\
    Angular resolution                & 0.1--1.7 & 0.1--3.3 \\
    Dimuon mass resolution            & 0.1--1.0 & 0.1--1.5 \\
    Fitting procedure                 & 0.1--3.2 & 0.4--25 \\
    Background distribution           & 0.1--7.2 & 0.1--29 \\[\cmsTabSkip]
    Total systematic uncertainty      & 1.6--7.5 & 4.4--39 \\
    \end{scotch}
    \label{tab:sysAFBFH}
  \end{table}

The finite size of the simulated event samples can affect the accuracy
of the efficiency determination. To estimate the uncertainty, 200
alternative efficiency functions are created by varying the parameters
of the signal efficiency function $\epsilon(\cos\theta_{\ell})$ within
their uncertainties. These alternative efficiencies are independently
used to fit the data. The standard deviations of the resulting $A_{\mathrm{FB}}$
and $F_{\PH}$ fit values are taken as their systematic uncertainties from
this source. The systematic uncertainty due to the efficiency
description is estimated by changing the modeling of
$\epsilon(\cos\theta_{\ell})$. The fit to $\epsilon(\cos\theta_{\ell})$ is
modified from a sixth-order polynomial to
the product of a Gaussian function and a
sixth-order polynomial, where the Gaussian function parameters are the
fit results from $\epsilon_{\text{acc}}$, and the sixth-order polynomial
parameters are the fit results from $\epsilon_{\text{reco}}$. The
differences in the results of $A_{\mathrm{FB}}$ and $F_{\PH}$ are used as the
systematic uncertainties.

The simulated signal sample is used to evaluate the effects of any
simulation mismodeling.
The difference in the fitted values of $A_{\mathrm{FB}}$ and $F_{\PH}$
between a simulated sample at the generator
level without the detector simulation and reconstruction steps, and the standard
signal simulation sample is assigned as the systematic
uncertainty. The specific parametrization of the function used to fit
the backgrounds can cause the results to change.
To evaluate the effect of fitting the background $\cos\theta_{\ell}$
distribution, the degrees of the polynomials used to describe the
angular shapes of the combinatorial background are decreased by
one. After fitting with the alternative background parametrization,
the differences in the $A_{\mathrm{FB}}$ and $F_{\PH}$ results
are taken as the systematic uncertainties from the background
parametrization model. The systematic uncertainties coming from the
experimental resolution in $\cos\theta_{\ell}$ and $q^2$ are estimated
by comparing the values of $A_{\mathrm{FB}}$ and $F_{\PH}$
obtained from the reconstructed MC events with those found using the
generated values of $\cos\theta_{\ell}$ and $q^2$ in the fit.

An estimate of the systematic uncertainty from the fitting procedure
is calculated using two different methods. In the first method, we
divide the large simulated signal sample into multiple subsamples,
each with a size similar to that of the data.
The difference between the average of the fitted values of $A_{\mathrm{FB}}$
and $F_{\PH}$ from the subsamples and the fitted value
from the full sample is taken as an estimate of the systematic
uncertainty from the modeling of the signal.
In the second method, we generate many pseudo-experiments
in which each of the mass and $\cos\theta_{\ell}$ distributions
are obtained from combining a signal and background distribution.
The signal distribution is obtained by selecting signal events from
the simulated sample, with the number of events determined by
the fit to the data. The background distribution is obtained
from sampling a parent distribution that comes from subtracting
the fitted signal distributions from the data.
The mean value of the differences from these pseudo-experiments
and the measurements from the reconstruction-level simulated signal
sample is taken as an estimate of the fitting uncertainty due to the
presence of background. The estimates from the two methods are then
added in quadrature to obtain the overall systematic uncertainty from
the fitting procedure.

In some $q^2$ ranges there are visible structures in the background
$\cos\theta_\ell$ distributions, as seen in
Fig.~\ref{fig:FHAFBdatacosl}. We have investigated many possible
contributions to these structures, and none of them has been
identified. This uncertainty is estimated using the ``second'' method
from the fitting procedure systematic uncertainty calculation, with the
$\cos\theta_{\ell}$ distribution for the background obtained
separately from the lower- and higher-mass sideband regions,
5.10--5.21 and 5.35--5.60\GeV. The larger of the two differences
between these alternative fits and the nominal fit is taken as the
systematic uncertainty from fitting the background $\cos\theta_{\ell}$
distribution.

The systematic uncertainties are estimated for each $q^2$ range independently.
As the systematic uncertainty sources are considered to be independent,
they are added in quadrature to obtain the total systematic
uncertainties, as shown in the last row of Table~\ref{tab:sysAFBFH}.

\section{Results}
\label{sec:Results}

To evaluate the statistical uncertainties, the 68.3\% confidence level
intervals on $A_{\mathrm{FB}}$ and $F_{\PH}$ are estimated using the profiled
Feldman--Cousins technique~\cite{Feldman_Cousins}. When estimating the
uncertainty in $A_{\mathrm{FB}}$ and $F_{\PH}$, the other variable is treated as a
nuisance parameter and profiled. A large number of pseudo-experiments are
generated with the maximum-likelihood estimate of the nuisance
parameter. The correlation between the two variables is ignored by
setting the confidence interval after using this profiling method.
The systematic and statistical uncertainties are added in
quadrature to obtain the total uncertainty.

\begin{figure}[!htb]
    \centering
	\includegraphics[width=0.49\textwidth]{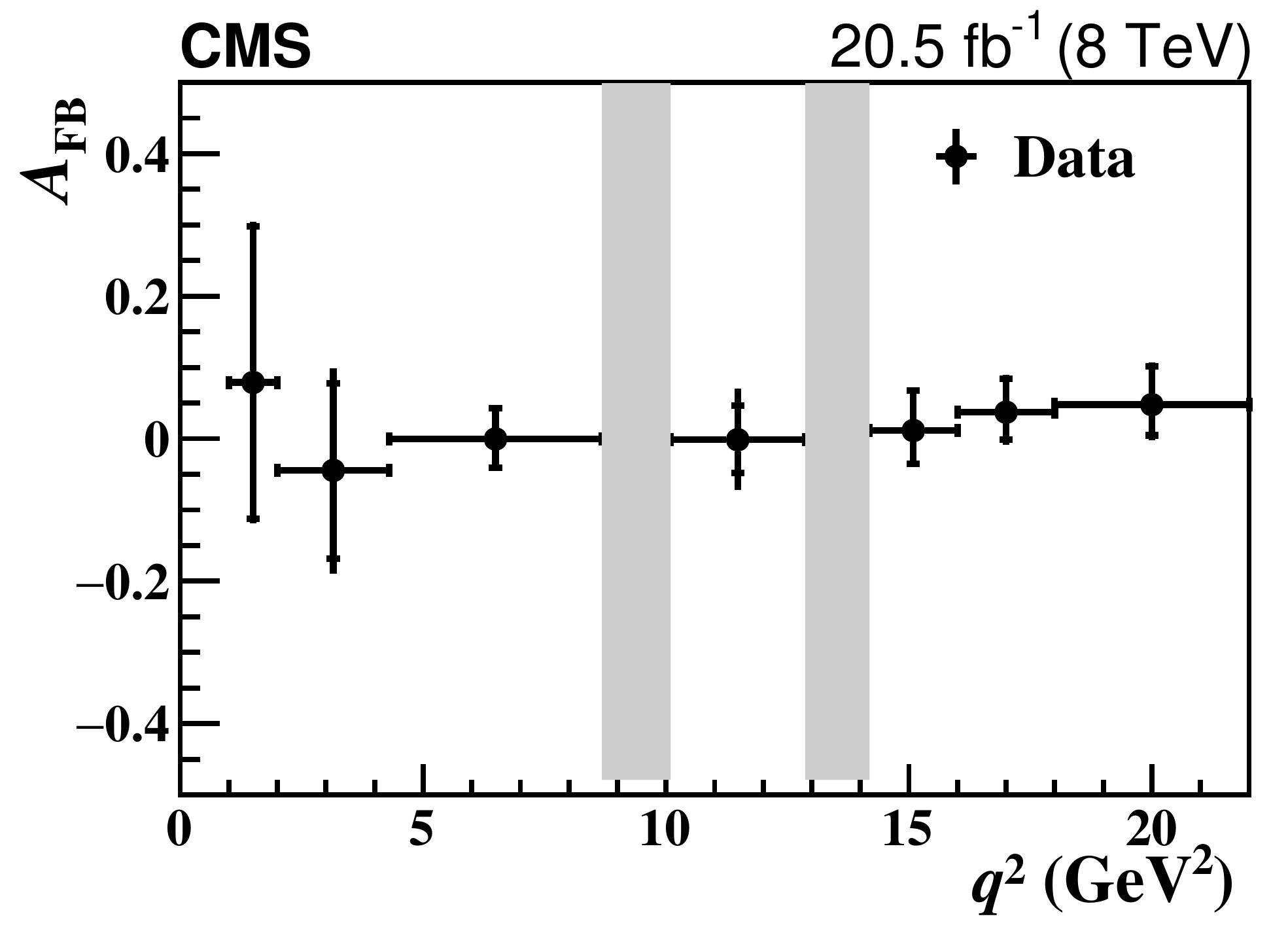}
    \includegraphics[width=0.49\textwidth]{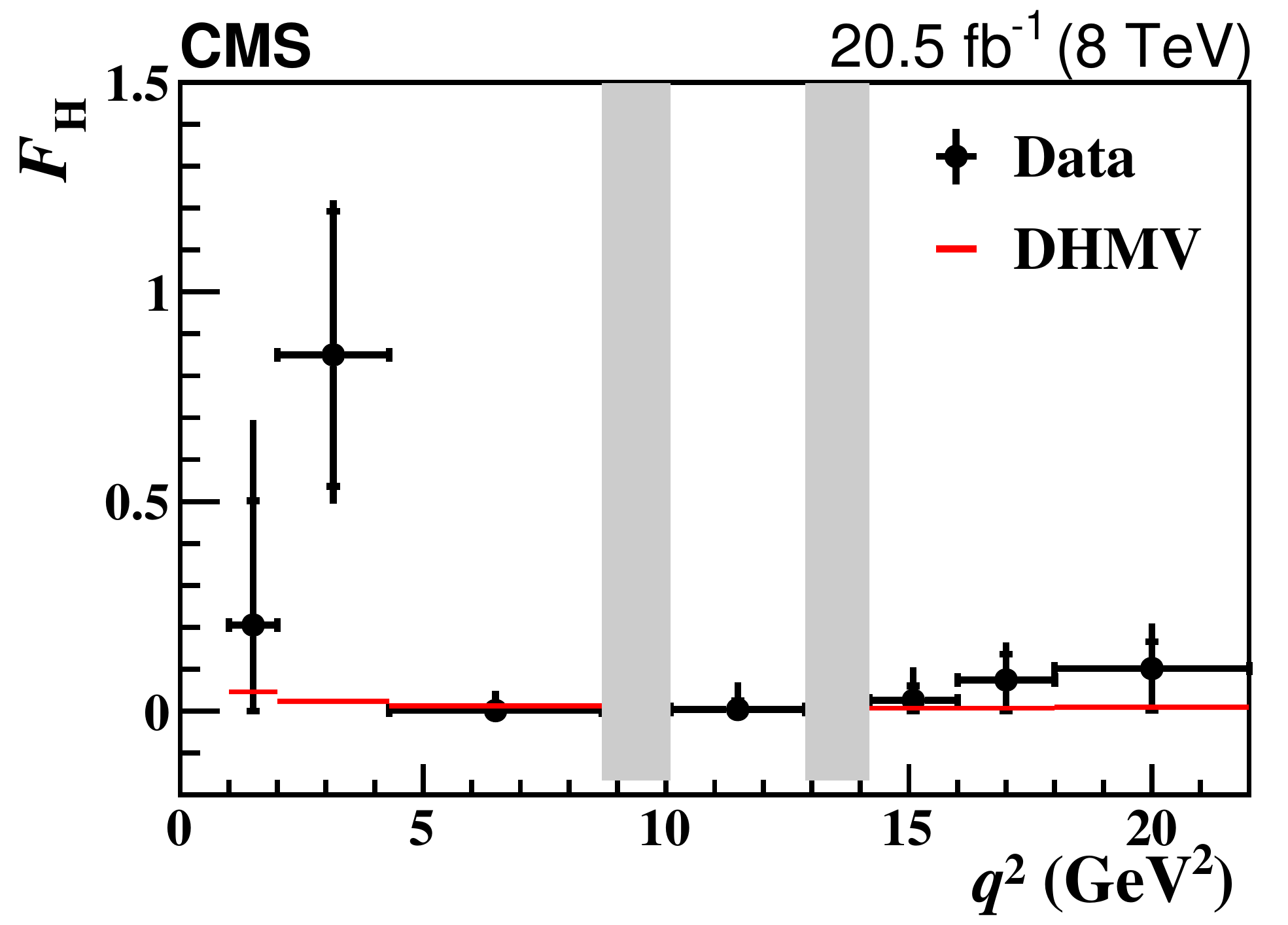}
    \caption{Results of the $A_{\mathrm{FB}}$ (\cmsLeft) and
      $F_{\PH}$ (\cmsRight) measurements in ranges of $q^2$. The
      statistical uncertainties are shown by the inner vertical bars,
      while the outer vertical bars give the total uncertainties. The
      horizontal bars show the $q^2$ range widths. The vertical
      shaded regions are 8.68--10.09 and 12.86--14.18$\GeV^2$,
      corresponding to the \JPsi- and $\Pgy$-dominated control
      regions, respectively. The horizontal lines in the right plot
      show the DHMV SM theoretical
      predictions~\cite{Descotes-Genon:2014uoa,Descotes-Genon:2015uva},
      whose uncertainties are smaller than the line width. }
    \label{fig:resultFHAFB}
\end{figure}

The measured values of $A_{\mathrm{FB}}$ and $F_{\PH}$ for each
$q^2$ range are shown in Fig.~\ref{fig:resultFHAFB}. The numerical
results are summarized in Table~\ref{tab:resFHAFB}, including the two
special $q^2$ ranges. The measured values of $A_{\mathrm{FB}}$ are
consistent with the SM expectation of no
asymmetry. Table~\ref{tab:resFHAFB} also includes three SM predictions
for $F_{\PH}$ with different input parameters and different
handling of higher-order corrections, one of which is also shown in
Fig.~\ref{fig:resultFHAFB}. There is generally good agreement between the
predictions and our results, as well as between our results and
previous measurements~\cite{Babar2006,Belle2009,CDF2012,LHCb2012An,LHCb2014An}.

\begin{table*} [!htb]
  \centering
  \topcaption{Results of the fit for each $q^2$ range,
    together with several SM predictions. The inclusive
	$q^2=$ 1.00--22.00$\GeV^2$ range in the bottom line does not include
    events from the \JPsi and $\Pgy$ resonance regions. The signal
    yield $Y_{\mathrm{S}}$ is given, along with its statistical
    uncertainty. The measured values of $A_{\mathrm{FB}}$ and
    $F_{\PH}$ are presented, where the first uncertainties are
    statistical and the second are systematic. The fifth column is a
    theoretical prediction by C. Bobeth et
    al.~\cite{Theory:2007,Bobeth:2011nj} using the EOS
    package~\cite{EOS} with the form factors from
    Refs.~\cite{Bouchard:2013pna,Ball2005,Khodjamirian2010}. The sixth
    column is the calculation from S. Descotes-Genon et al.~(DHMV)
    based on
    Refs.~\cite{Descotes-Genon:2014uoa,Descotes-Genon:2015uva}. The
    last column is the prediction using the FLAVIO
    package~\cite{Straub:2018kue} with the form factors from
    Ref.~\cite{flavio_1}. Only the central values of the theoretical
    predictions are shown, since their uncertainties are insignificant
    compared to those in the measurements.}
	\cmsTable{
    \renewcommand\arraystretch{1.3}
	\begin{scotch}{ccccccc}
	   $q^2$ (${\GeVns}^2$) & $Y_{\mathrm{S}}$ & $A_{\mathrm{FB}}$ & $F_{\PH}$ & ${F_{\PH}}$(EOS) & ${F_{\PH}}$(DHMV) & ${F_{\PH}}$(FLAVIO) \\
	   \hline
	   1.00--2.00   & $~~169 \pm 22$ & $~~~0.08~^{+0.22}_{-0.19} \pm 0.05$ & $0.21~^{+0.29}_{-0.21} \pm 0.39$ &  0.047  &  0.046  &  0.045  \\
	   2.00--4.30   & $~~331 \pm 32$ & $  -0.04~^{+0.12}_{-0.12} \pm 0.07$ & $0.85~^{+0.34}_{-0.31} \pm 0.14$ &  0.024  &  0.023  &  0.022  \\
	   4.30--8.68   & $~~785 \pm 42$ & $~~~0.00~^{+0.04}_{-0.04} \pm 0.02$ & $0.01~^{+0.02}_{-0.01} \pm 0.04$ & \NA     &  0.012  &  0.011  \\
	   10.09--12.86 & $~~365 \pm 29$ & $~~~0.00~^{+0.05}_{-0.05} \pm 0.05$ & $0.01~^{+0.02}_{-0.01} \pm 0.06$ & \NA     & \NA     & \NA     \\
	   14.18--16.00 & $~~215 \pm 19$ & $~~~0.01~^{+0.06}_{-0.05} \pm 0.02$ & $0.03~^{+0.03}_{-0.03} \pm 0.07$ &  0.007  &  0.007  &  0.006  \\
	   16.00--18.00 & $~~262 \pm 21$ & $~~~0.04~^{+0.05}_{-0.04} \pm 0.03$ & $0.07~^{+0.06}_{-0.07} \pm 0.07$ &  0.007  &  0.007  &  0.006  \\
	   18.00--22.00 & $~~226 \pm 20$ & $~~~0.05~^{+0.05}_{-0.04} \pm 0.02$ & $0.10~^{+0.06}_{-0.10} \pm 0.09$ &  0.008  &  0.009  &  0.008  \\[\cmsTabSkip]
	   1.00--6.00   & $~~778 \pm 47$ & $  -0.14~^{+0.07}_{-0.06} \pm 0.03$ & $0.38~^{+0.17}_{-0.21} \pm 0.09$ &  0.025  &  0.025  &  0.020  \\[\cmsTabSkip]
     ~~1.00--22.00  & $ 2286 \pm 73$ & $~~~0.00~^{+0.02}_{-0.02} \pm 0.03$ & $0.01~^{+0.01}_{-0.01} \pm 0.06$ & \NA     & \NA     & \NA     \\
    \end{scotch}
	}
	\label{tab:resFHAFB}
\end{table*}

\section{Summary}

An angular analysis of the decay \Sig has been performed using a data
sample of proton-proton collisions corresponding to an integrated
luminosity of $20.5\fbinv$ recorded with the CMS detector at
$\sqrt{s} = 8\TeV$. The forward-backward asymmetry $A_{\mathrm{FB}}$
of the muon system and the contribution $F_{\PH}$ of the
pseudoscalar, scalar, and tensor amplitudes to the decay width are
measured as a function of the dimuon mass squared. The results are
consistent with previous measurements, and are also compatible with
three different standard model predictions.

\begin{acknowledgments}
  We congratulate our colleagues in the CERN accelerator departments
  for the excellent performance of the LHC and thank the technical and
  administrative staffs at CERN and at other CMS institutes for their
  contributions to the success of the CMS effort. In addition, we
  gratefully acknowledge the computing centers and personnel of the
  Worldwide LHC Computing Grid for delivering so effectively the
  computing infrastructure essential to our analyses. Finally, we
  acknowledge the enduring support for the construction and operation
  of the LHC and the CMS detector provided by the following funding
  agencies: BMWFW and FWF (Austria); FNRS and FWO (Belgium); CNPq,
  CAPES, FAPERJ, and FAPESP (Brazil); MES (Bulgaria); CERN; CAS, MoST,
  and NSFC (China); COLCIENCIAS (Colombia); MSES and CSF (Croatia);
  RPF (Cyprus); SENESCYT (Ecuador); MoER, ERC IUT, and ERDF (Estonia);
  Academy of Finland, MEC, and HIP (Finland); CEA and CNRS/IN2P3
  (France); BMBF, DFG, and HGF (Germany); GSRT (Greece); NKFIA
  (Hungary); DAE and DST (India); IPM (Iran); SFI (Ireland); INFN
  (Italy); MSIP and NRF (Republic of Korea); LAS (Lithuania); MOE and
  UM (Malaysia); BUAP, CINVESTAV, CONACYT, LNS, SEP, and UASLP-FAI
  (Mexico); MBIE (New Zealand); PAEC (Pakistan); MSHE and NSC
  (Poland); FCT (Portugal); JINR (Dubna); MON, RosAtom, RAS and RFBR
  (Russia); MESTD (Serbia); SEIDI, CPAN, PCTI and FEDER (Spain); Swiss
  Funding Agencies (Switzerland); MST (Taipei); ThEPCenter, IPST,
  STAR, and NSTDA (Thailand); TUBITAK and TAEK (Turkey); NASU and SFFR
  (Ukraine); STFC (United Kingdom); DOE and NSF (USA).

  \hyphenation{Rachada-pisek} Individuals have received support from
  the Marie-Curie program and the European Research Council and
  Horizon 2020 Grant, contract No. 675440 (European Union); the
  Leventis Foundation; the A. P. Sloan Foundation; the Alexander von
  Humboldt Foundation; the Belgian Federal Science Policy Office; the
  Fonds pour la Formation \`a la Recherche dans l'Industrie et dans
  l'Agriculture (FRIA-Belgium); the Agentschap voor Innovatie door
  Wetenschap en Technologie (IWT-Belgium); the F.R.S.-FNRS and FWO
  (Belgium) under the ``Excellence of Science - EOS" - be.h project
  n. 30820817; the Ministry of Education, Youth and Sports (MEYS) of
  the Czech Republic; the Lend\"ulet (``Momentum") Program and the
  J\'anos Bolyai Research Scholarship of the Hungarian Academy of
  Sciences, the New National Excellence Program \'UNKP, the NKFIA
  research grants 123842, 123959, 124845, 124850 and 125105 (Hungary);
  the Council of Science and Industrial Research, India; the HOMING
  PLUS program of the Foundation for Polish Science, cofinanced from
  European Union, Regional Development Fund, the Mobility Plus program
  of the Ministry of Science and Higher Education, the National
  Science Center (Poland), contracts Harmonia 2014/14/M/ST2/00428,
  Opus 2014/13/B/ST2/02543, 2014/15/B/ST2/03998, and
  2015/19/B/ST2/02861, Sonata-bis 2012/07/E/ST2/01406; the National
  Priorities Research Program by Qatar National Research Fund; the
  Programa Estatal de Fomento de la Investigaci{\'o}n Cient{\'i}fica y
  T{\'e}cnica de Excelencia Mar\'{\i}a de Maeztu, grant MDM-2015-0509
  and the Programa Severo Ochoa del Principado de Asturias; the Thalis
  and Aristeia programs cofinanced by EU-ESF and the Greek NSRF; the
  Rachadapisek Sompot Fund for Postdoctoral Fellowship, Chulalongkorn
  University and the Chulalongkorn Academic into Its 2nd Century
  Project Advancement Project (Thailand); the Welch Foundation,
  contract C-1845; and the Weston Havens Foundation (USA).
\end{acknowledgments}

\bibliography{auto_generated}
\cleardoublepage \appendix\section{The CMS Collaboration \label{app:collab}}\begin{sloppypar}\hyphenpenalty=5000\widowpenalty=500\clubpenalty=5000\vskip\cmsinstskip
\textbf{Yerevan Physics Institute, Yerevan, Armenia}\\*[0pt]
A.M.~Sirunyan, A.~Tumasyan
\vskip\cmsinstskip
\textbf{Institut f\"{u}r Hochenergiephysik, Wien, Austria}\\*[0pt]
W.~Adam, F.~Ambrogi, E.~Asilar, T.~Bergauer, J.~Brandstetter, E.~Brondolin, M.~Dragicevic, J.~Er\"{o}, A.~Escalante~Del~Valle, M.~Flechl, R.~Fr\"{u}hwirth\cmsAuthorMark{1}, V.M.~Ghete, J.~Hrubec, M.~Jeitler\cmsAuthorMark{1}, N.~Krammer, I.~Kr\"{a}tschmer, D.~Liko, T.~Madlener, I.~Mikulec, N.~Rad, H.~Rohringer, J.~Schieck\cmsAuthorMark{1}, R.~Sch\"{o}fbeck, M.~Spanring, D.~Spitzbart, A.~Taurok, W.~Waltenberger, J.~Wittmann, C.-E.~Wulz\cmsAuthorMark{1}, M.~Zarucki
% ...
\vskip\cmsinstskip
% ...
1:  Also at Vienna University of Technology, Vienna, Austria\\
% ...
\end{sloppypar}
\end{document}